\documentclass[aps,prb,amsmath,amssymb,bibnotes,reprint,longbibliography,superscriptaddress]{revtex4-1}

\usepackage{t1enc}
\usepackage[utf8]{inputenc}
\usepackage{amsmath}

\usepackage{bbold}
\usepackage{graphicx}
\usepackage{siunitx}
\usepackage{hyperref}
\usepackage{braket}
\usepackage{cases}
\usepackage{cancel}
\usepackage[overload]{empheq}

\usepackage{xcolor}

\newcommand{\noteandras}[1]{\textcolor{black}{#1}}

\begin{document}

\title{Dephasing of 
Majorana qubits due to quasistatic disorder}

\author{P\'{e}ter Boross}
\affiliation{Institute for Solid State Physics and Optics, Wigner Research Centre for Physics, H-1525 Budapest P.O. Box 49, Hungary}
\affiliation{Department of Theoretical Physics and MTA-BME Exotic Quantum Phases Research Group,
Budapest University of Technology and Economics, H-1111 Budapest, Hungary}

\author{Andr\'as P\'alyi}
\affiliation{Department of Theoretical Physics and MTA-BME Exotic Quantum Phases Research Group,
Budapest University of Technology and Economics, H-1111 Budapest, Hungary}

\date{\today}

\begin{abstract}
Quantum bits based
on Majorana zero modes are expected to
be robust against certain noise types, and
hence provide a quantum computing 
platform that is superior to conventional
qubits. 
This robustness is not complete though:
imperfections can still lead to 
qubit decoherence and hence to
information loss.
In this work, we theoretically 
study Majorana-qubit dephasing in a minimal
model: in a Kitaev chain with quasistatic disorder.
Our approach, based on numerics as
well as first-order
non-degenerate perturbation theory, 
provides a conceptually simple
physical picture 
and predicts Gaussian dephasing.
We show that, as system parameters
are varied, the dephasing rate due to disorder
oscillates out-of-phase with respect to the
oscillating \noteandras{Majorana} splitting of the clean system.
\noteandras{In our model, first-order dephasing sweet spots are absent
if disorder is uncorrelated. 
We describe the crossover between uncorrelated and highly correlated
disorder, and show that dephasing measurements can be used
to characterize the disorder correlation length.}
We expect that our results will be utilized for 
the design and interpretation of future
Majorana-qubit experiments.
\end{abstract}

\maketitle

\section{Introduction}
\label{sec:introduction}

Theoretical proposals\cite{OregPRL2010,LutchynPRL2010,SauPRB2010} suggest that Majorana zero modes (MZMs) can be engineered in quasi-one-dimensional 
semiconducting-superconducting 
hybrid systems \cite{MourikScience2012,PradaNatRevPhys2020}.
The past decade has witnessed intense
experimental activities to establish MZMs\cite{MourikScience2012,DasNatPhys2012,DengNanoLett2012,FinckPRL2013,ChurchillPRB2013,DengSciRep2014,AlbrechtNature2016,ShermanNatNano2017,DengSciAdv2016,SuominenPRL2017,NichelePRL2017,GulNatNano2018,DengPRB2018,GrivninNatComm2019,VaitiekenasScience2020,VaitiekenasNatPhys2021}.
It is expected that MZMs could serve
as building blocks in experiments 
demonstrating topologically protected 
quantum memories, quantum dynamics, or
even quantum computing 
\cite{AliceaNatPhys2011,AliceaRPP2012,HasslerNJP2011,vanHeckNJP2012,HyartPRB2013,
AasenPRX2016,KarzigPRB2017,TutschkuPRB2020}.
In that context, understanding the
decoherence of Majorana qubits\cite{BrouwerPRL2011,GoldsteinPRB2011,SchmidtPRB2012,BudichPRB2012,RainisPRB2012,PedrocchiPRL2015,AasenPRX2016,KnappPRB2018dephasing,AseevPRB2018,BauerSciPost2018,LaiPRB2018,AseevPRB2019,MishmashPRB2020}
is an important task.

The minimal model 
hosting MZMs is the Kitaev
chain \cite{Kitaev2001}.
It can be used to 
describe the dephasing process
of a Majorana qubit. The ground
state of a finite-length topological
Kitaev chain hosts two MZMs at the
two ends of the chain, implying that
the ground state is approximately 
twofold degenerate, with one ground
state being of even fermion parity
and the other being of odd fermion parity.
In a chain with a finite length, 
a small energy splitting $\varepsilon_0$ separates the
two ground states. 
If random components, such as disorder\cite{BrouwerPRL2011,HegdePRB2016},
are incorporated in 
the model, then the splitting becomes a random
variable.
To encode a single qubit with MZMs, two
wires and hence four MZMs are needed \cite{Leijnse_2012}.
In such a two-wire Majorana qubit, the random splittings in the two
wires add up to a random Larmor frequency of the
qubit, leading to qubit dephasing. 

In this work, we theoretically study
dephasing of Majorana qubits in 
the presence of slow charge noise.
A key target 
in topological quantum computing is the
experimental
demonstration of a
topologically protected quantum memory based
on MZMs,
hence it is imperative to understand
the potential sources of qubit decoherence, 
to assess future device functionality
and provide optimization guidelines.
Furthermore, qubit dephasing measurement
is an established tool to reveal
the noise structure of the qubit's environment
\cite{CywinskiPRB2008,BylanderNatPhys2011,DialPRL2013,YonedaNatNano2018};
understanding dephasing is important 
for that application, too. 
In our work, we focus on the model 
of quasistatic disorder \cite{BorossNanoTech2016,TosiNatComm2017,BorossPRB2018,BoterPRB2020,SzechenyiPRB2020,DerakhshanPhysRevApp2020},
a minimal model of slow (low-frequency)
charge noise or 
$1/f$ noise \cite{ShnirmanPhysScri2002,CywinskiPRB2008, DialPRL2013,FreemanAPL2016,YonedaNatNano2018,HetenyiPRB2019, CywinskiPRB2020,MishmashPRB2020},
which has been a very important 
source of qubit dephasing 
both in semi- and superconductor environments.

Naturally, Majorana qubit dephasing
due to weak and slow (quasistatic) charge noise is 
determined by the probability distribution
of the splitting, see, e.g., our 
section \ref{sec:dephasing}. 
First we study that splitting distribution 
\noteandras{for uncorrelated disorder}
by both numerical and analytical methods,
and show that it is Gaussian for weak
disorder.
We argue that this result is consistent with the 
log-normal splitting-envelope distribution
found by Ref.~\onlinecite{BrouwerPRL2011} in our Appendix~\ref{appendix:compareBrouwer}.

Having the splitting distribution at hand, we use it to characterize 
the dephasing of a Majorana qubit subject
to weak quasistatic disorder.
In simple models, the time dependence
of qubit dephasing often 
follows a Gaussian function\cite{AasenPRX2016,HansonRMP2007}.
Here we show that this is also the case 
for the qubit studied here.
\noteandras{Our key results for the spatially uncorrelated disorder model}
 are that 
(i) we provide
an analytical formula for the dephasing susceptibility [see Eq.~\eqref{eq:approxdephasingsusc}]
and the dephasing time [see Eq.~\eqref{eq:T2star}], 
(ii) we reveal an out-of-phase oscillation 
between the splitting of the clean
system and the dephasing susceptibility to 
disorder 
[see Fig.~\ref{fig:splittingcleananddirty}b and Fig.~\ref{fig:splittingnumana}a-b],
(iii) and we highlight the absence of dephasing sweet spots
in our model (see Sec.~\ref{sec:dephasing}).
\noteandras{
Finally, we show that the spatial correlation length of the disorder
has a strong impact on a dephasing experiment.
As a consequence, we expect that in future Majorana-qubit 
experiments, measuring the dephasing time as function
of control parameters (e.g., chemical potential) will provide
information about the spatial structure of noise.}

The rest of the paper is organized as follows. 
In section \ref{sec:numerics},
we show numerical results for the splitting
distribution of the disordered chain, 
highlighting the 
Gaussian distribution of the splitting, 
and the out-of-phase oscillation between
the clean splitting and the
splitting susceptibility to disorder.
In section \ref{sec:continuum}, we
use the continuum version of the Kitaev chain, 
together with mode matching and first-order
perturbation theory, to establish the
semi-analytical description of the splitting
distribution, and to derive approximate
analytical results for that.  
\noteandras{Furthermore we compare the results of the two models.}
In section \ref{sec:dephasing}, we
relate the splitting distribution and 
the dephasing dynamics of a Majorana qubit
based on two Kitaev chains. 
\noteandras{In section \ref{sec:spatialdistribution}, we show 
that the parameter-dependence of the dephasing time 
is sensitive to the correlation length of the disorder.}
We discuss implications and follow-up ideas
in section \ref{sec:discussion},
and conclude in section \ref{sec:conclusions}.

\begin{figure}
    \includegraphics[width=8.6cm]{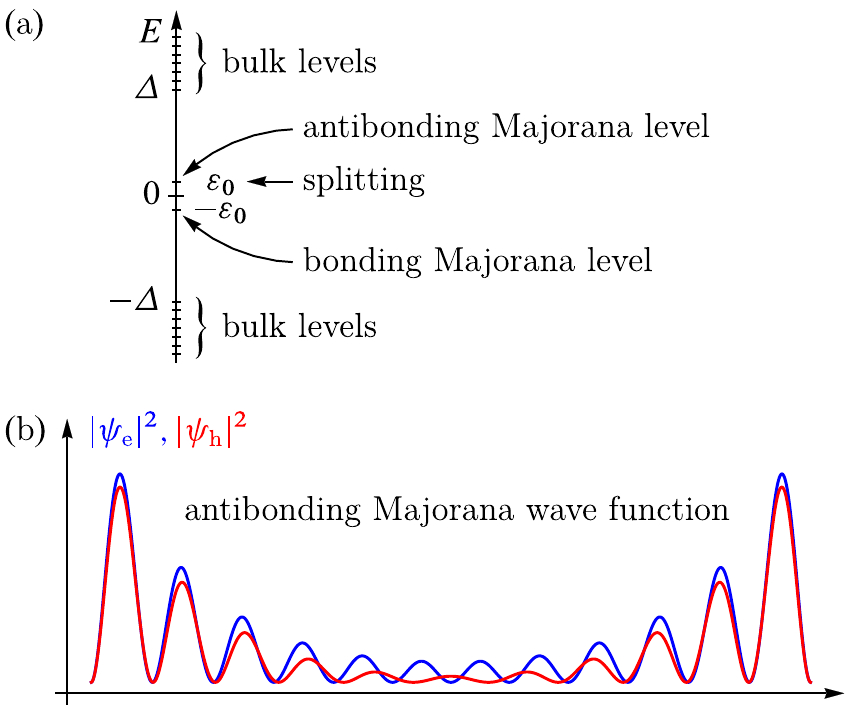}
    \centering
    \caption{Spectrum and Majorana wave functions
    in a topological superconductor wire. 
    (a) Schematic spectrum of the Bogoliubov-de Gennes matrix.
    (b) Electron and hole components of the antibonding Majorana wave function}
    \label{fig:wfandspectrum}
\end{figure}

\section{Disorder-induced splitting distribution in the Kitaev chain}
\label{sec:numerics}

We use the Kitaev-chain tight-binding model\cite{Kitaev2001} to numerically 
investigate Majorana qubit dephasing. 
In this section, we numerically determine the 
disorder-induced
distribution of the signful splitting (see definition below), 
using the Kitaev chain.
We anticipate that this distribution is
Gaussian for weak quasistatic disorder
(see below within this section for details),
and that the
Majorana qubit dephasing time $T_2^*$ in a two-chain setup
is determined by the standard deviation $\sigma_{\epsilon_0}$ of the signful splitting
$\epsilon_0$
[see Eq.~\eqref{eq:T2star} in 
section \ref{sec:dephasing}].

The Hamiltonian of a finite-length 
Kitaev chain in real space reads
\cite{Kitaev2001}
\begin{align}
\label{eq:kitaevchain}
    H_\text{K} =& -\sum_{n=1}^{N}
    \left(\mu_\text{K}+\delta\mu_n^{(\text{K})}\right)
    {c_n^\dagger c_n^{\vphantom\dagger}} -t\sum_{n=1}^{N-1}{\left(c_n^\dagger c_{n+1}^{\vphantom\dagger}+\text{h.c.}\right)} \nonumber \\
    &-\Delta_\text{K}\sum_{n=1}^{N-1}{\left(c_n^{\vphantom\dagger} c_{n+1}^{\vphantom\dagger}+\text{h.c.}\right)},
\end{align}
where $c_{n}^\dagger$ and $c_{n}^{\vphantom\dagger}$ are the electron creation and annihilation operator on site $n$, respectively, $t$ is the hopping amplitude, $\mu_\text{K}$ is the chemical potential, $\Delta_\text{K}$ is the superconducting pair
potential, and $N$ is the number of sites.
We model disorder as a random on-site potential,
independent on each site, 
drawn from Gaussian distribution with zero mean and standard deviation $\sigma_\mu$, that
is,     
$\delta\mu_n^{(\text{K})}\sim\mathcal{N}(0,\sigma_\mu)$.
For a discussion of the relation between this model and
disorder in real samples, see section \ref{sec:discussion}.

\begin{figure*}
\includegraphics[width=17.8cm]{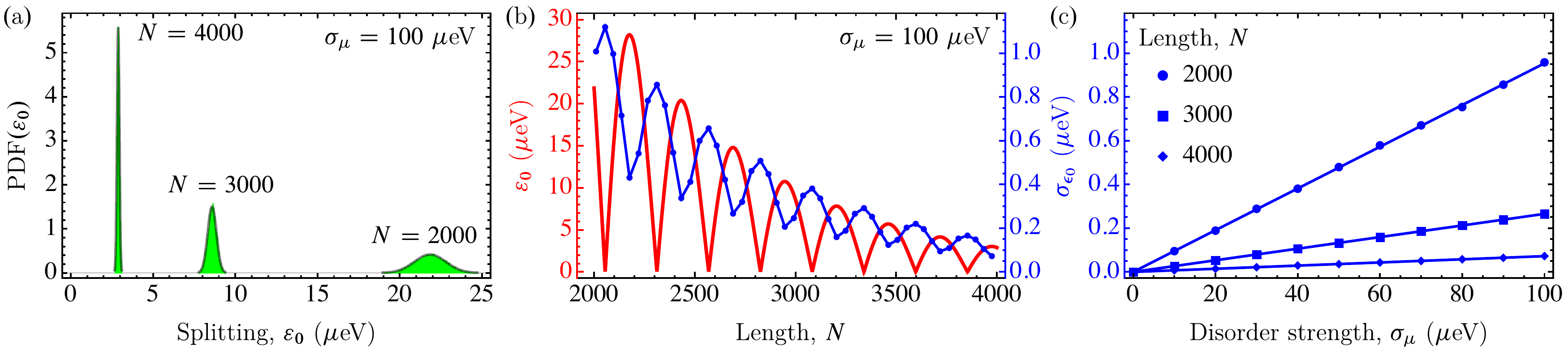}
\centering
\caption{Splitting, its probability distribution and its standard deviation from the Kitaev chain model. (a) Numerically obtained probability density functions (pdfs) of the splitting for three different lengths in disordered system. Gray lines are fitted Gaussian pdfs. (b) Splitting of the clean system (red line) and standard deviation of the signful splitting (blue points) are shown as a function of chain length, for the disorder strength $\sigma_\mu = 100\text{ $\mu$eV}$. Out-of-phase oscillation can be observed between the splitting and its standard deviation. (c) Standard deviation of the signful splitting is shown as a function of the strength of the on-site disorder for three different lengths. The dependence on the disorder strength is linear for the shown range. Results for disordered systems are calculated using 10000 realizations.}
\label{fig:splittingcleananddirty}
\end{figure*}

We obtain the splitting
$\varepsilon_0$ using the Bogoliubov-de Gennes (BdG)
transformation\cite{ScheurerPRB2013}, i.e., 
by numerically finding the smallest positive eigenvalue of the corresponding real-space $2N\times2N$ BdG Hamiltonian\footnote{To obtain the smallest positive eigenvalue, we apply the \texttt{Eigenvalues} function of Wolfram Mathematica as \texttt{Eigenvalues[HBdG, 1, Method -> \{"Arnoldi", "Shift" -> 0\}]} in the version 12.0.0.0}.
We will calculate the splitting
$\varepsilon_0$ for a clean system, i.e., in the absence of any disorder, as
well as for random on-site disorder realizations. 
In the latter case, $\varepsilon_0$ becomes a random variable -- 
with Gaussian distribution
for weak disorder, as 
shown in Fig.~\ref{fig:splittingcleananddirty}a
and discussed below.

A clean Kitaev chain has a splitting
that decreases in an oscillatory
fashion as the chain length
is increased \cite{DasSarmaPRB2012,PientkaNJP2013,ThakurathiJphys2015,GiladPRB2015}.
This is shown in Fig.~\ref{fig:splittingcleananddirty}b,  
where we
plot the numerically calculated
length dependence of $\varepsilon_0$
(red solid line)
for a parameter set shown in the 
`Kitaev chain' section of 
Table \ref{tab:parameters}.

\begin{table}
\begin{tabular}{l  c  c}
Parameter/scale & Notation & Value \\
\hline \hline
Continuum model \\ \hline
Effective mass (of InAs) & $m$ & $0.023 m_\text{e}$ \\
Chemical potential & $\mu_\text{C}$ & $1\text{ meV}$ \\
Superconducting gap & $\Delta_\text{C}$ & $200\text{ $\mu$eV}$ \\
\hline
Kitaev chain \\
\hline
Normal hopping amplitude & $t$ &  $6.62\text{ eV}$  \\
Lattice constant & $a$ & $0.5\text{ nm}$ \\
Chemical potential & $\mu_\text{K}$ & $-13.3\text{ eV}$ \\
Superconducting pairing potential & $\Delta_\text{K}$ & $8.14\text{ meV}$ \\
\hline
Length scales \\
\hline
Fermi-wavelength & $\lambda_\text{F}$ & $511a$ \\
Fermi-wavenumber & $k_\text{F}$ & $0.0123/a$ \\
Superconductor coherence length & $\xi$ & $814a$ \\
Inverse coherence lenght & $\kappa$ & $0.00123/a$
\end{tabular}
    \caption{Parameter values used in the numerical and analytical calculations.
	\label{tab:parameters}}
\end{table}

Now we introduce disorder and study the
splitting distribution.
Figure \ref{fig:splittingcleananddirty}a shows three examples of that
distribution, for a fixed disorder strength
$\sigma_\mu = 100\, \mu$eV, 
for three different chain lengths
$N = 2000, \, 3000, \, 4000$.
The figure clearly shows a Gaussian character
for all three distributions.
Furthermore, the figure also shows the trend
that both
the mean and the standard deviation of these
distributions decrease as the chain length
increases.

Figure \ref{fig:splittingcleananddirty}c shows a more systematic analysis
of the length- and disorder-strength
dependence of the 
standard deviation $\sigma_{\epsilon_0}$
of the signful splitting $\epsilon_0$.
Note the difference between the \emph{signful splitting}
$\epsilon_0$ and the \emph{splitting}
$\varepsilon_0$.
The signful splitting is defined by $\epsilon_0 \equiv \epsilon_\text{o}-\epsilon_\text{e}$, where $\epsilon_\text{o}$ ($\epsilon_\text{e}$) is the energy of the odd (even) ground state. 
We have defined the splitting (see Fig.~\ref{fig:wfandspectrum}a) as the absolute value of the signful splitting, i.e. $\varepsilon_0 \equiv |\epsilon_0|$.
The distinction between
$\varepsilon_0$ and $\epsilon_0$ 
is motivated by the observation
that the dephasing dynamics is related to 
the signful splitting
$\epsilon_0$, 
see Eq.~\eqref{eq:T2star}.

For all lengths displayed in 
Fig.~\ref{fig:splittingcleananddirty}c, 
the standard
deviation $\sigma_{\epsilon_0}$
of the signful splitting
shows a clear linear dependence
on the disorder strength $\sigma_\mu$. 
This linear dependence motivates the
definition of the dimensionless
\emph{dephasing susceptibility
to disorder}, $\chi = \sigma_{\epsilon_0}/{\sigma_\mu}$.
In Eq.~\eqref{eq:dephasingsusc}, 
we will provide an approximate
analytical formula for this susceptibility.

The Gaussian character of the 
splitting distribution, and the linear
dependence of the splitting standard deviation
on the disorder strength can be qualitatively
understood in three steps. 
We briefly summarize these here, and 
will use these considerations in 
the next section in our quantitative 
derivations.

(1)~The bonding and antibonding
Majorana levels, see Fig.~\ref{fig:wfandspectrum}a,
are particle-hole
symmetric partners of each other.
This implies that disorder (or any 
other perturbation) 
can not couple them directly.

(2)~Therefore, there is no need to use
degenerate or quasi-degenerate perturbation
theory to describe the leading-order effect
of disorder on the energy levels.
It is sufficient to do 
first-order non-degenerate perturbation theory
for, say, the antibonding level.
This explains the linear dependence 
of the splitting standard deviation on the
disorder strength.

(3)~The first-order perturbative description
implies that the 
first-order 
energy correction $\delta \varepsilon_0^{(1)}$ 
due to disorder in our model (independent random
on-site energies) is a sum of many independent
random variables for long chains, $N\gg 1$,
and hence the central limit theorem 
ensures the Gaussian character of that energy 
correction.

Finally, we point out an out-of-phase oscillation effect
between the standard deviation $\sigma_{\epsilon_0}$
of the signful splitting and the clean splitting $\varepsilon_0$.
In Fig.~\ref{fig:splittingcleananddirty}b, 
the blue points show the length dependence
of the standard deviation $\sigma_{\epsilon_0}$
of the signful splitting,
for the disorder strength $\sigma_\mu = 100\text{ $\mu$eV}$. 
Note that the y axis for these blue points
is the right y axis which is also colored
blue. 
Fig.~\ref{fig:splittingcleananddirty}b shows that
the splitting standard deviation
$\sigma_{\epsilon_0}$ oscillates and decays
as the length increases, similarly
to the splitting of the clean system.
However, there is an out-of-phase oscillation between the splitting and its disorder-induced 
standard deviation: e.g., the standard deviation has a maximum wherever the splitting reaches zero.

Note that with our numerical approach, 
it is straightforward to estimate the standard 
deviation  $\sigma_{\varepsilon_0}$ of the splitting;
however, Majorana-qubit dephasing is determined by 
the standard deviation $\sigma_{\epsilon_0}$ of the 
signful splitting [Eq.~\eqref{eq:T2star}]. 
To estimate the latter, we do the following. 
If the expectation value of the splitting $\varepsilon_0$ 
is much larger than its standard deviation, then
$\sigma_{\epsilon_0} \approx \sigma_{\varepsilon_0}$, 
hence we use the splitting statistics to estimate
$\sigma_{\epsilon_0}$. 
If the above condition does not hold, 
then we convert the statistics of the splitting 
to the statistics of the signful splitting,
and from the latter we estimate $\sigma_{\varepsilon_0}$, 
as described in Appendix \ref{appendix:signfulstd}.

\section{Splitting in the continuum version of the Kitaev chain}
\label{sec:continuum}

Numerical computation of the splitting from the Kitaev chain model or other tight-binding models can be computationally expensive for larger system size. To establish a more efficient calculational tool, and to enable analytical results for the splitting absolute value and its standard deviation, here we study the continuum version of the Kitaev chain.
These analytical results serve also as a benchmark 
against which the numerical results can be checked.

First, we use mode matching to
obtain the BdG wave function of the
quasi-zero-energy mode
in a clean (disorder-free) wire.\cite{ThakurathiJphys2015} 
Second, we use this wave function and
first-order 
non-degenerate perturbation theory to 
determine the standard deviation 
$\sigma_{\epsilon_0}$ 
of the splitting. 

\subsection{Splitting and antibonding Majorana wave function in a clean wire}

The continuum model has the following
momentum-space Hamiltonian\cite{BrouwerPRL2011}:
\begin{equation}
\label{eq:continuum}
    \mathcal{H}_\text{C}(k) = \left(\frac{\hbar^2k^2}{2m}-\mu_\text{C}\right)\sigma_z - \Delta_\text{C}' \hbar k \sigma_x,
\end{equation}
where $m$ is the effective mass, $\mu_\text{C}$ is the chemical potential, and  $\sigma_x$ and $\sigma_z$ are Pauli matrices acting in Nambu space. The index C stands for `continuum'. For future use, we define
\begin{subequations}
    \begin{align}
        \label{eq:k_F}
        k_\text{F} &= \sqrt{2 m \mu_\text{C}}/\hbar, \\
        \label{eq:v_F}
        v_\text{F} &= \hbar k_\text{F}/m, \\
        \label{eq:Delta_C}
        \Delta_\text{C} &=\Delta_\text{C}' \hbar k_\text{F}, \\
        \label{eq:xi}
        \xi &= \hbar v_\text{F}/\Delta_\text{C},
    \end{align}
\end{subequations}
where $k_F$ is the Fermi wave number, $v_\text{F}$ is the Fermi-velocity, $\Delta_\text{C}$ is the superconducting gap and $\xi$ is the superconductor coherence length.
We will describe a finite-length wire
with length $L$ and hard-wall boundary conditions.
The relation of this Hamiltonian 
and the Kitaev-chain Hamiltonian is detailed
in Appendix~\ref{appendix:connecttwomodel}.

We use mode-matching to determine
the Majorana antibonding state and its
energy (the splitting).
The first step is 
to establish the evanescent modes close
to zero energy in a homogeneous system.
With that aim, we insert 
the standard plane-wave ansatz
to the BdG equation 
$\mathcal{H}_\text{C}(-i \partial_x) \psi(x)
= \varepsilon \psi(x)$
defined by Eq.~\eqref{eq:continuum}.

It is
straightforward to show that this approach
yields four evanescent solutions for energies $0\leq\varepsilon<\sqrt{\Delta_\text{C}^2-\left(\Delta_\text{C}^2/2\mu_\text{C}\right)^2}$, with complex wave numbers $k_1= K + i \kappa$, $k_2= -K + i \kappa$, $k_3= K - i \kappa$ and $k_4= -K - i \kappa$.
Here
\begin{subequations}
    \label{eq:wavenumbers}
	\begin{align}
		K &= \frac{1}{\hbar}\sqrt{m\left(\sqrt{\mu_\text{C}^2-\varepsilon^2}+\mu_\text{C}-\frac{\Delta_\text{C}^2}{2\mu_\text{C}}\right)}, \\
		\kappa &= \frac{1}{\hbar}\sqrt{m\left(\sqrt{\mu_\text{C}^2-\varepsilon^2}-\mu_\text{C}+\frac{\Delta_\text{C}^2}{2\mu_\text{C}}\right)}.
	\end{align}
\end{subequations}
Furthermore, $K$ and $\kappa$ are positive numbers for $0\leq\varepsilon<\sqrt{\Delta_\text{C}^2-\left(\Delta_\text{C}^2/2\mu_\text{C}\right)^2}$. 
The corresponding non-normalized wave functions have the form 
\begin{equation}
    \label{eq:wavefunctionk}
    \psi_{k_i}(x) = \begin{pmatrix} u_{k_i} \\ v_{k_i} \end{pmatrix} e^{i k_i x} =  \begin{pmatrix}
				\Delta_\text{C}' \hbar {k_i} \\
				\frac{\hbar^2 {k_i}^2}{2m}-\mu_\text{C}-\varepsilon
				\end{pmatrix} e^{i k_i x},
\end{equation}
where $u_{k_i}$ and $v_{k_i}$ represent the electron and hole components of the wave function in the momentum space.

The antibonding Majorana wave function
must be a linear superposition of the 
four evanescent modes at a given energy:
\begin{equation}
    \label{eq:wavefunctionbasic}
	\psi(x)= \begin{pmatrix}\psi_\text{e}(x) \\ \psi_\text{h}(x) \end{pmatrix} = \sum_{i=1}^{4}{\alpha_i \begin{pmatrix} u_{k_i} \\ v_{k_i} \end{pmatrix} e^{i k_i x}},
\end{equation}
where $\psi_\text{e}(x)$ and $\psi_\text{h}(x)$ are the electron and hole components of the Majorana bound state, furthermore $\alpha_i$-s are complex coefficients. 
This wave function $\psi(x)$ 
must satify the hard-wall boundary
conditions:
\begin{equation}
    \label{eq:boundaryconditions}
    \psi(0)=\psi(L) = \begin{pmatrix}0 \\0 \end{pmatrix}.
\end{equation}
This condition is fulfilled by
coefficient vectors satisfying 
the following homogeneous
linear set of equations: 
\begin{equation}
	\label{eq:nullspace}
	\mathcal{M} \begin{pmatrix} \alpha_1 \\ \alpha_2 \\ \alpha_3 \\ \alpha_4 \end{pmatrix} = \begin{pmatrix} 0 \\ 0 \\ 0 \\ 0 \end{pmatrix},
\end{equation}
where the $\varepsilon$-dependent matrix
$\mathcal{M}$ is defined as
\begin{equation}
	\mathcal{M} = \begin{pmatrix}
		u_{k_1} & u_{k_2} & u_{k_3} & u_{k_4} \\
		v_{k_1} & v_{k_2} & v_{k_3} & v_{k_4} \\
		u_{k_1} e^{i k_1 L} & u_{k_2} e^{i k_2 L} & u_{k_3} e^{i k_3 L} & u_{k_4} e^{i k_4 L}\\
		v_{k_1} e^{i k_1 L} & v_{k_2} e^{i k_2 L} & v_{k_3} e^{i k_3 L} & v_{k_4} e^{i k_4 L}
	\end{pmatrix}.
\end{equation}

As follows from Eq.~\eqref{eq:nullspace},
for a given length $L$, the condition
\begin{equation}
\label{eq:det}
	\det{\mathcal{(M(\varepsilon))}}=0
\end{equation}
gives the energy $\varepsilon_0$
of the antibonding Majorana state. 
In general, Eq.~\eqref{eq:det} leads a 
transcendental equation, which can be solved numerically: $\varepsilon_{0,\text{num}}$.
Power-series expansion of $\det{\mathcal{(M)}}$ in $\varepsilon$ up to second order provides an analytical solution that in the limit of $L\gg 1/k_\text{F},\xi$ reads
\begin{equation}
    \label{eq:generalsplitting}
	\varepsilon_0(L) \approx 2\Delta_\text{C} k_\text{F} e^{-L/\xi} \left|\frac{\sin\left(\sqrt{k_\text{F}^2-1/\xi^2} L\right)}{\sqrt{k_\text{F}^2-1/\xi^2}}\right|,
\end{equation}
where we use $K|_{\varepsilon=0} = \sqrt{k_\text{F}^2-1/\xi^2}$ and $\kappa|_{\varepsilon=0} = 1/\xi$.

Depending on the relative magnitude of $k_\text{F}$ and $1/\xi$,
from Eq.~\eqref{eq:generalsplitting}
we obtain 
\begin{widetext}
    \begin{subequations}
        \label{eq:pientkageneralized}
        \begin{align}[left={\varepsilon_0(L) \approx \empheqlbrace}]
            & \Delta_\text{C} \frac{2k_\text{F}}{\sqrt{k_\text{F}^2-1/\xi^2}} e^{-L/\xi}\left|\sin\left(\sqrt{k_\text{F}^2-1/\xi^2} L\right)\right|, &\text{if }k_\text{F}>1/\xi, \\
            & \Delta_\text{C} \frac{k_\text{F}}{\sqrt{1/\xi^2-k_\text{F}^2}} e^{-\left(1/\xi-\sqrt{1/\xi^2-k_\text{F}^2}\right)L}, &\text{if }k_\text{F}<1/\xi.
        \end{align}
    \end{subequations}
\end{widetext}
If $k_\text{F}>1/\xi$ (i.e.~when $\mu_\text{C} >\Delta_\text{C}/2$), the splitting has an oscillatory part, but if $k_\text{F}<1/\xi$, the splitting decreases purely exponentially as the length increases. For a physically feasible parameter set shown in Table \ref{tab:parameters}, including a 
chemical potential (e.g., set by a gate voltage)
$\mu_\text{C} = 1\, \text{meV}$,
we obtain $k_\text{F}>1/\xi$.
This is the case we focus on from now on. 
To reach $k_\text{F} < 1/ \xi$, 
the chemical potential needs to be 
suppressed as $0 < \mu_\text{C} < 0.1\, \text{meV}$; we do not
treat this case here.

Our result (\ref{eq:pientkageneralized}a) is in fact
a generalization of an earlier result,
see below Eq.~(5) in Ref.~\onlinecite{PientkaNJP2013} \noteandras{and Eq.~(18) in Ref.~\onlinecite{ZengPRB2019}.}
The only difference is the appearance of $\sqrt{k_\text{F}^2-1/\xi^2}$
in our result. The earlier result can be obtained
by taking the limit $k_\text{F} \gg 1/\xi$ of our
formula (\ref{eq:pientkageneralized}a), i.e., 
by applying the approximation 
$\sqrt{k_\text{F}^2-1/\xi^2} \approx k_\text{F}$.

Next, we describe the antibonding Majorana wave function.
To simplify the description, 
we
utilize the symmetries of the setup.
The clean system has inversion symmetry.
The corresponding operator has the from $\Pi = \pi\,\otimes\,\sigma_z$, where $\pi$ is the inversion with respect to the point $x=L/2$,
acting in real space, and $\sigma_z$ acts 
in Nambu space. 
Inversion symmetry, together
with the assumption that 
the antibonding Majorana energy level 
is non-degenerate, implies that
\begin{equation}
    \label{eq:inversion}
    \begin{pmatrix} \psi_\text{e}(x) \\ \psi_\text{h}(x)  \end{pmatrix} = \begin{pmatrix} \pm\psi_\text{e}(L-x) \\ \mp\psi_\text{h}(L-x)  \end{pmatrix}.
\end{equation}

The Hamiltonian Eq.~\eqref{eq:continuum} also has bosonic time-reversal symmetry with the operator $T=(\mathbb{1}\,\otimes\,\sigma_z)\mathcal{K}$, where $\mathcal{K}$ is the complex conjugation, and it fulfills the relation $T^2=1$. Time-reversal symmetry restricts the form of the non-degenerate 
energy eigenstate as
\begin{equation}
    \label{eq:timereversal0}
    T \begin{pmatrix} \psi_\text{e}(x) \\ \psi_\text{h}(x)  \end{pmatrix} = \begin{pmatrix} \psi_\text{e}^*(x) \\ -\psi_\text{h}^*(x)  \end{pmatrix} = e^{i\varphi} \begin{pmatrix} \psi_\text{e}(x) \\ \psi_\text{h}(x)  \end{pmatrix},
\end{equation}
where $\varphi$ depends on the global phase of the wave function. 
For concreteness, we fix this global phase such that
$\varphi = 0$.
Given an eigenstate $\psi$ with an arbitrary global phase, eigenstate with $\varphi = 0$ is obtained as 
$\psi(x)+T\psi(x)$.
This choice $\varphi = 0$ leads to
\begin{subequations}
    \label{eq:timereversal}
    \begin{align}
        \text{Im}[\psi_\text{e}(x)] &= 0, \\
        \text{Re}[\psi_\text{h}(x)] &= 0.
    \end{align}
\end{subequations}

Equations~\eqref{eq:boundaryconditions}, \eqref{eq:inversion} and \eqref{eq:timereversal} constrain the form of the wave function:
\begin{widetext}
\begin{equation}
    \label{eq:wavefunction}
    \psi(x) =
		\begin{pmatrix}
			A_\text{e} \left\{e^{-\kappa x} \sin\left(K x - \phi_\text{e} \right) + p e^{-\kappa (L-x)} \sin\left[K (L-x) - \phi_\text{e} \right]\right\}\\
			i A_\text{h} \left\{e^{-\kappa x} \sin\left(K x - \phi_\text{h} \right) - p e^{-\kappa (L-x)} \sin\left[K (L-x) - \phi_\text{h} \right]\right\}
		\end{pmatrix},
\end{equation}
\end{widetext}
where $A_\text{e}$ and $A_\text{h}$ are normalization factors,
\begin{subequations}
    \label{eq:phaseshiftp}
    \begin{align}
        \phi_\text{e} &= \arctan\left[\frac{p e^{-\kappa L}\sin(K L)}{1 +p e^{-\kappa L}\cos(K L)}\right], \\
        \phi_\text{h} &= \arctan\left[\frac{-p e^{-\kappa L}\sin(K L)}{1 -p e^{-\kappa L}\cos(K L)}\right]
    \end{align}
\end{subequations}
are phases, and $p= +1$ ($p= -1$)
corresponds to the behavior under inversion, that
is, to the upper (lower) sign in Eq.~\eqref{eq:inversion}.

In the limit $L \gg \xi > 1/k_\text{F}$, the following
approximations can be made:
\begin{subequations}
    \label{eq:longlimitapprox}
	\begin{align}
	    p &= \text{sign}\left[\sin\left(\sqrt{k_\text{F}^2-1/\xi^2} L\right)\right], \\
		\phi_\text{e} &= - \phi_\text{h} = \phi \approx e^{-L/\xi} \left|\sin\left(\sqrt{k_\text{F}^2-1/\xi^2} L\right)\right|, \\
		A_\text{e} &= A_\text{h} = A\approx \frac{1}{\sqrt{\frac{\xi}{2}\left(1-\frac{\cos(\phi/2)}{k_F^2\xi^2}\right)}}.
	\end{align}
\end{subequations}
We note that $p$ changes sign where the splitting vanishes.

To obtain  Eq.~(\ref{eq:longlimitapprox}a), we compare the
wave function in Eqs.~\eqref{eq:wavefunctionbasic} and
in \eqref{eq:wavefunction}, yielding
\begin{equation}
    \label{eq:phaseshiftau}
    \phi_\text{e}=\arctan{\left[\frac{\alpha_1 u_{k_1}+\alpha_2 u_{k_2}}{i\left(\alpha_2 u_{k_2}-\alpha_1 u_{k_1}\right)}\right]}.
\end{equation}
The coefficient vector $(\alpha_1,\alpha_2,\alpha_3,\alpha_4)^\intercal$ is the nullspace of the matrix $\mathcal{M}$, which we find 
analytically by Gauss elimination.
Comparing Eqs.~(\ref{eq:phaseshiftp}a) and \eqref{eq:phaseshiftau} 
up to leading order in $e^{-L/\xi}$, we find
Eq.~(\ref{eq:longlimitapprox}a). 
Furthermore, we find the approximate formula for the phases 
in Eq.~(\ref{eq:longlimitapprox}b) using
Eqs.~\eqref{eq:phaseshiftp}, by taking leading-order approximation
in $e^{-L/\xi}$,
and utilizing Eq.~(\ref{eq:longlimitapprox}a). 
To obtain Eq.~(\ref{eq:longlimitapprox}c), we assumed that the electron and hole character of the wave function has exactly equal probability in the limit of $L \gg \xi > 1/k_\text{F}$, which results in $A_\text{e}=A_\text{h}$.
We derived Eq.~(\ref{eq:longlimitapprox}c) from the norm of wave function in Eq.~\eqref{eq:wavefunction} by taking the limit for $L\to\infty$. We will use Eqs.~\eqref{eq:longlimitapprox} to derive an approximate analytical formula for the dephasing susceptibility to disorder shown in Eq.~\eqref{eq:approxdephasingsusc}.

\subsection{Standard deviation of the splitting}

Now we describe the broadening of the splitting distribution due to on-site disorder.
The full Hamiltonian of the disordered system can be written as
\begin{equation}
\label{eq:contdisorder}
    H_\text{C} = \mathcal{H}_\text{C}(-i\hbar\partial_x) + H_\text{dis},
\end{equation}
where $H_\text{dis} = \delta\mu_\text{C}(x)\sigma_z$ is the disorder Hamiltonian, representing
disorder in the chemical potential.
We model disorder as a collection of 
potential steps, where the lengths of the
steps are equal and denoted by $a_\text{dis}$:
\begin{equation}
    \label{eq:contdisorder}
	\delta\mu_\text{C}(x) = \sum_{i=1}^{N_\text{dis}}{\delta\mu_i^{(\text{C})} \Xi_i(x)},
\end{equation}
where
\begin{equation}
	\noteandras{\Xi_i(x)} = \begin{cases} 
      1 & i-1\leq x/a_\text{dis}< i, \\
      0 & \text{otherwise}.
   \end{cases}
\end{equation}
This model is a natural analog of the disorder
model we used in the Kitaev chain, with the
identification $a = a_\text{dis}$, where $a$ is the lattice constant of the Kitaev chain.

We regard disorder as a perturbation, and calculate the first order energy shift. 
Naively, one should do degenerate perturbation theory, since the antibonding and bonding Majorana
energies are close to each other.
However, disorder does not couple them,
hence non-degenerate perturbation theory is 
sufficient.
The proof of this is as follows.

Due to the particle-hole symmetry of the BdG
Hamiltonian: $\braket{\psi|H_\text{dis}|P\psi} = \braket{P\psi|H_\text{dis}|\psi} = 0$, where $\ket{\psi}$ and $\ket{P\psi}$ are the positive and negative energy solution of the BdG Hamiltonian, and $P=(\mathbb{1}\,\otimes\,\sigma_x)\mathcal{K}$ is the operator of the particle-hole symmetry. This can be seen by
\begin{multline}
     \braket{\psi|H|P\psi} 
     =  -\braket{\psi|PH|\psi} = \\
    = -\braket{P\psi|H|\psi}^*  = -\braket{H\psi|P\psi} =-\braket{\psi|HP\psi},
\end{multline}
where the first equation is implied by the fact that $H$ anticommutes with $P$, 
the second equation is the consequence of the anti-unitary property of $P$, 
the third equation is obtained by flipping the scalar product, and the fourth equation
is implied by $H$ being Hermitian.

Applying the 
relation 
$\braket{\psi|H_\text{dis}|P\psi} = 0$
to the antibonding $\ket{\psi}$ and
bonding $\ket{P\psi}$ Majorana wave functions, 
we conclude that they are uncoupled
and therefore the first-order disorder-induced 
shift of the signful splitting is
$\delta \varepsilon_0^{(1)} = 
\braket{\psi| H_\text{dis} |\psi}$.
Using Eqs.~\eqref{eq:wavefunctionbasic} and
\eqref{eq:contdisorder},
this shift can be written as
\begin{equation}
    \label{eq:disorderME}
	\braket{\psi| H_\text{dis} |\psi} = \sum_{i=1}^{N_\text{dis}}\delta\mu_i^{(\text{C})} \Theta_i,
\end{equation}
where
\begin{equation}
	\noteandras{\Theta_i}=\int_{(i-1)a_\text{dis}}^{i a_\text{dis}}{\left[\left|\psi_\text{e}(x)\right|^2 -\left|\psi_\text{h}(x)\right|^2 \right]\text{d}x}.
\end{equation}

In analogy with our disorder model in the Kitaev
chain, 
discussed in section \ref{sec:numerics},
we assume independence and normal distribution 
for the chemical potential disorder, which we denote as $\delta\mu_i^{(\text{C})}\sim\mathcal{N}(0,\sigma_\mu)$, where $\sigma_\mu$ is the disorder strength, 
and $\delta\mu_i^{(\text{C})}$-s are independent of each other. 
From Eq.~\eqref{eq:disorderME}, we conclude that the disorder matrix element also follows Gaussian distribution:
\begin{equation}
\label{eq:gaussian}
	\braket{\psi|H_\text{dis}|\psi} \sim \mathcal{N}\left(0,\sigma_{\epsilon_0}\right),
\end{equation}
where $\sigma_{\epsilon_0}=\sigma_\mu\sqrt{\sum_{j=1}^{N_\text{dis}}\Theta_j^2}$ the standard deviation of the distribution of the
signful splitting.

Let us suppose that $\left|\psi_\text{e}(x)\right|^2 -\left|\psi_\text{h}(x)\right|^2$ varies slowly on the scale of $a_\text{dis}$.
This implies
\begin{align}
	\sum_{i=1}^{N_\text{dis}}\Theta_i^2 &\approx \sum_{i=1}^{N_\text{dis}}{a_\text{dis}^2 \left[\left|\psi_\text{e}(i a_\text{dis})\right|^2 -\left|\psi_\text{h}(i a_\text{dis})\right|^2\right]^2} \nonumber \\
	&\approx a_\text{dis}\int_0^L{\left[\left|\psi_\text{e}(x)\right|^2 -\left|\psi_\text{h}(x)\right|^2\right]^2\text{d}x}.
\end{align}
Therefore, the dephasing susceptibility to disorder is obtained
as
\begin{equation}
    \label{eq:dephasingsusc}
    \chi \equiv \frac{\sigma_{\epsilon_0}}{\sigma_\mu} = \sqrt{a_\text{dis}\int_0^L{\left[\left|\psi_\text{e}(x)\right|^2 -\left|\psi_\text{h}(x)\right|^2\right]^2\text{d}x}}.
\end{equation}
By solving Eq.~\eqref{eq:nullspace} numerically, 
we obtain the values $\alpha_{i,\text{num}}$ 
of $\alpha_{i}$.
Substituting these numerical values $\alpha_{i,\text{num}}$ and $\varepsilon_{0,\text{num}}$ into Eqs.~\eqref{eq:wavenumbers} and \eqref{eq:wavefunction}, 
we obtain the semi-analytical wave funciton
$\psi_\text{e}(x)$ and $\psi_\text{h}(x)$.
After normalization, Eq.~\eqref{eq:dephasingsusc} can be performed. Results, shown in Fig.~\ref{fig:splittingnumana}b as 'exact', are
discussed below.

As an alternative to the above semi-analytical
approach, an approximate analytical formula can be obtained by substituting the form of the wave function in Eqs.~\eqref{eq:wavefunction} into Eq.~\eqref{eq:dephasingsusc} using Eqs.~\eqref{eq:longlimitapprox}. After the integration over $x$, and taking series expansion in $\kappa e^{-\kappa L}$, the dephasing susceptibility to disorder in limit of $L \gg \xi > 1/k_\text{F}$ can be written as
\begin{multline}
    \label{eq:approxdephasingsusc}
    \chi =\sqrt{\frac{a_\text{dis}}{2\xi}} e^{-L/\xi} \times \\
    \times\sqrt{\frac{8L}{\xi}-3+\left(\frac{4L}{\xi}+3\right)\cos\left(2\sqrt{k_\text{F}^2-1/\xi^2}L\right)}.
\end{multline}

Equation \eqref{eq:approxdephasingsusc}
is the key result of our work. 
It reveals that 
the dephasing susceptibility (and hence the dephasing rate)
as a function of system parameters
exhibits oscillations that are out-of-phase 
with the oscillations of the clean splitting given in 
Eq.~\eqref{eq:pientkageneralized}. 
This is apparent as Eq.~\eqref{eq:pientkageneralized}
contains a sine whereas 
Eq.~\eqref{eq:approxdephasingsusc} contains a cosine.

Furthermore, Eq.~\eqref{eq:approxdephasingsusc}
also suggests the absence of dephasing sweet spots in this
setting: the long expression below the square root
in Eq.~\eqref{eq:approxdephasingsusc} is always positive
due to the condition $L \gg \xi$.

In conclusion, we have
described a semi-analytical procedure,
and an approximate analytical procedure, 
to estimate the disorder-induced broadening
of the distribution of the signful splitting 
in a continuum model of a 1D topological 
superconductor.

\subsection{Comparing the results of the two models}
\label{sec:compare}

\begin{figure*}
\includegraphics[width=17.8cm]{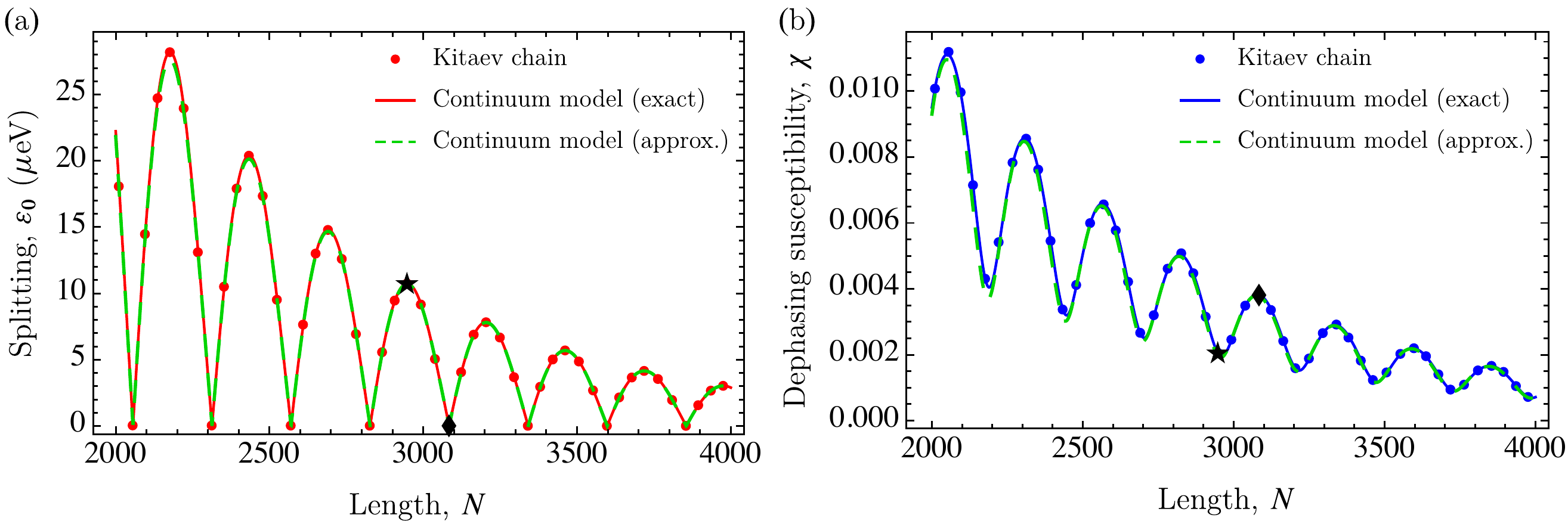}
\centering
\caption{Comparison of continuum-model (lines) and Kitaev-model (dots) results for the splitting and its standard deviation. (a) Splitting as a function of length for the clean system. (b) Dephasing
susceptibility to disorder -- that is, the ratio
$\chi = \sigma_{\epsilon_0} / \sigma_\mu$
of the standard deviation of the splitting and the strength of the on-site disorder -- is shown as a function of the length.
See Table \ref{tab:parameters} for parameter values. 
In both panels, star and diamond denote two specific chain lengths, for which
the dephasing curves are shown in Fig.~\ref{fig:dephasingsummary}.}
\label{fig:splittingnumana}
\end{figure*}

Here, we show the correspondence 
of the Kitaev chain and continuum model 
results for the splitting $\varepsilon_0$,
the disorder-induced standard deviation $\sigma_{\epsilon_0}$, and the 
dephasing susceptibility $\chi$. 
\noteandras{In Appendix \ref{appendix:connecttwomodel}, we show how to connect the parameters of the continuum and discrete (Kitaev) models.}

In Fig.~\ref{fig:splittingnumana}a, we plot the splitting of the clean system as a function of the chain length. 
Red points show the numerical result from the Kitaev chain model, whereas the red solid line
shows the semi-analytical exact  result from the continuum model, obtained by solving Eq.~\eqref{eq:det} numerically. 
Dashed green line
shows the result of Eq.~(\ref{eq:pientkageneralized}a).
Parameter values are those listed in Table \ref{tab:parameters}. 
In Fig.~\ref{fig:splittingnumana}a, 
the Kitaev chain result (red points) and the exact
result from the continuum model (red solid line)
are indistinguishable. 
The analytical approximate result (green dashed line)
shows a slight deviation from the other two data sets
for short chain, 
but becomes indistinguishable from
those for long chains.

In Fig.~\ref{fig:splittingnumana}b, we plot the dephasing susceptibility,
that is, the ratio of splitting standard deviation 
$\sigma_{\epsilon_0}$
and the disorder strength $\sigma_\mu$, 
as the function of the chain length. 
The Kitaev model result (points) is obtained numerically, 
using 10000 random disorder realizations for each length.
Here again, the two models show satisfactory agreement.

\section{Signful splitting distribution and Majorana qubit dephasing}
\label{sec:dephasing}

In this section, we complete our primary
task, and describe the dephasing
dynamics of a Majorana qubit subject
to quasistatic disorder. 

\subsection{Noise model: quasistatic disorder}

\noteandras{Let us start this description by defining our noise model
of quasistatic disorder, and relating it to device physics. 
Electrical potential fluctuations are generically present in qubit 
devices, and often dominate qubit decoherence. 
In many experiments, this noise has been found to follow
a frequency-dependent power spectrum $S(f) \propto 1/f$. 
Due to dominance of the low-frequency component, 
one can refer to this type of noise as slow charge noise.}

\noteandras{In this work, we account for the most prominent feature 
of this noise, 
i.e., that it detunes the electrostatic potential felt by the
electrons in the Majorana wire. 
Regarding the spatial structure of the noise, we first focus on 
short-range correlations (Sec.~\ref{sec:dephasing}), but later we also 
describe Majorana
qubit dephasing as the spatial correlation length is varied 
(Sec.~\ref{sec:spatialdistribution}).}

\noteandras{Regarding the temporal structure of noise, we follow
numerous earlier works by applying the quasistatic approximation.
To define the quasistatic approximation, 
we first recall how a dephasing-time
experiment (\emph{Ramsey experiment}) is performed.
First, a balanced superposition of the two computational basis states, 
with a Bloch vector aligned with, say, the $x$ axis,  
is prepared.
Then, this state is allowed to evolve freely for a waiting time $\tau_w$
much shorter than the dephasing time. 
After time $\tau_w$, the qubit is measured in the $x$ basis. 
This is often called one `shot' of the experiment. 
This shot is repeated many ($N_\text{rep} \gg 1$) times
to gain statistics and eliminate shot noise,
and the whole sequence of $N_\text{rep}$ shots 
is repeated for $N_\tau \gg 1$ different, stepwise increasing
 values of the waiting time.
Typically, the largest $\tau_w$ value is a few times greater than
the dephasing time.}

\noteandras{As applied to this scheme, 
the quasistatic approximation of noise
is composed of two assumptions:
(i) for each run, the noise is considered time-independent, 
i.e., it is static disorder, and
(ii) for the $N_\text{rep}$ shots with a single waiting time, 
the different static disorder configurations acting 
during the different runs provide a good statistical coverage
of all disorder configurations.}

\subsection{Majorana qubit dephasing}

Consider a Majorana qubit encoded in 
two identical topological superconducting wires.
All parameters are assumed to be equal, including 
the disorder strength. 
The two wires are assumed to be decoupled
from each other (no tunneling between the
two wires). 
Restrict our attention to the globally even
ground state of this setup, which is spanned
by the basis states 
$\ket{0} \equiv \ket{e_1,e_2}$ and
$\ket{1} \equiv \ket{o_1,o_2}$, where
the names $e$ and $o$ refer to the
even and odd 
fermion parities of the corresponding 
states, and the 
indices 1 and 2 refer to the first and second wire,
respectively.

To perform a qubit dephasing experiment,
one usually creates an 
initial state $\ket{\psi_\text{i}}$ that
is a balanced superposition of the
two basis states, e.g., with a qubit
polarization vector along the x direction
\begin{equation}
\ket{\psi_\text{i}} = \frac{1}{\sqrt{2}}\left(
\ket{0} + \ket{1}
\right).
\end{equation}
The qubit polarization vector
(Bloch vector) for this state
is 
\begin{equation}
    \vec p \equiv \braket{\psi_\text{i} | \vec \sigma | \psi_\text{i} } = (1,0,0),
\end{equation}
where $\vec \sigma = (\sigma_x,\sigma_y,\sigma_z)$ 
is the vector of Pauli
matrices.
Note that the preparation of this initial state itself
can be corrupted by disorder, a complication that we disregard here.

After preparation, the relative phase 
between the two basis states evolves in time
due to the excess energy 
$\epsilon_0^{(1)}$ of $o_1$ with respect
to $e_1$ in wire 1,
and the excess energy 
$\epsilon_0^{(2)}$ of $o_2$ with respect
to $e_2$ in wire 2.
In particular, the time-dependent
wave function, up to an irrelevant
global phase, reads
\begin{eqnarray}
\ket{\psi(t)} = 
\frac{1}{\sqrt{2}}\left(
\ket{0}
+e^{-i (\epsilon_0^{(1)} + \epsilon_0^{(2)}) t /\hbar}\ket{1}
\right).
\end{eqnarray}
Then, the quasistatic assumption 
implies that on average, for a large number
of measurements, 
the qubit polarization vector evolves
in time as
\begin{equation}
\label{eq:polarization}
    \langle \vec p(t) \rangle \equiv
    \braket{\psi(t) | \vec \sigma| \psi(t)}
    = 
    \int d \epsilon\, \rho(\epsilon)
    \left(\begin{array}{c}
         \cos(\epsilon t /\hbar)\\
         -\sin(\epsilon t / \hbar)\\
         0
    \end{array}
    \right),
\end{equation}
where $\epsilon = \epsilon_0^{(1)} + \epsilon_0^{(2)}$
is the random qubit energy splitting, 
and $\rho(\epsilon)$ is its pdf.

\begin{figure*}
\includegraphics[width=17.8cm]{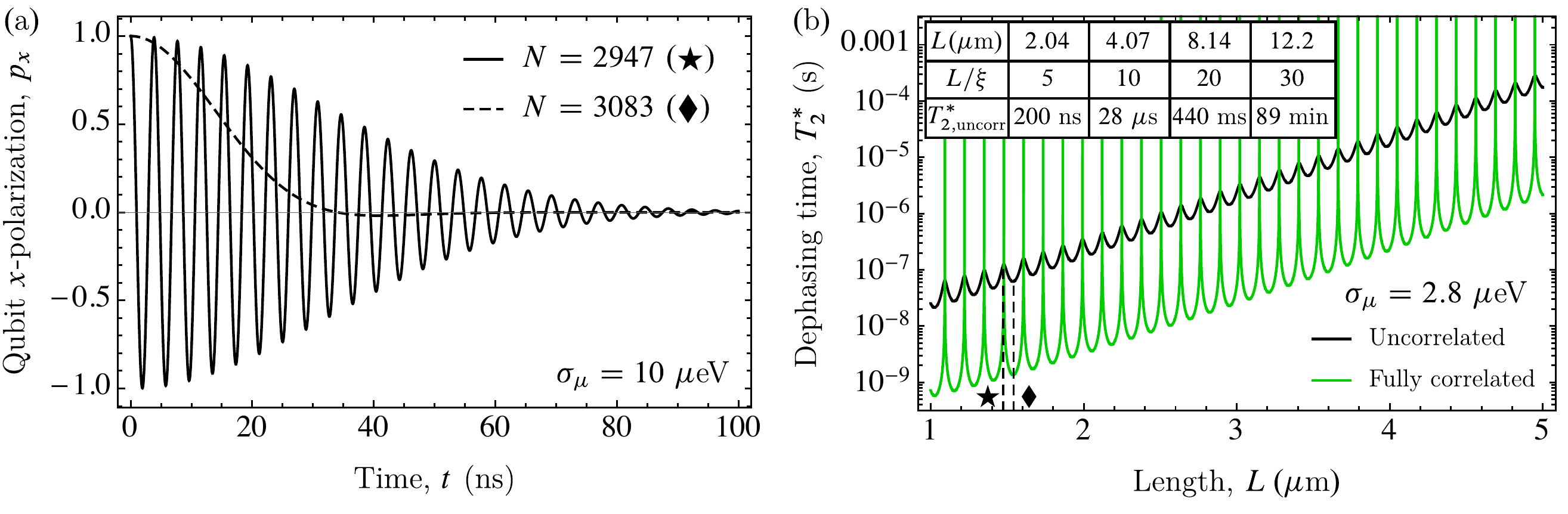}
\centering
\caption{Dephasing of a Majorana qubit due to quasistatic disorder.
(a) Dephasing curves. The $x$ component of the disorder-averaged polarization vector as a function of time for two different lengths (see star and diamond in Fig.~\ref{fig:splittingnumana}.) The envelopes of the curves follows Gaussian dephasing. The finiteness of the mean of the signful splitting is responsible for the oscillations of the solid line. The out-of-phase oscillation between the clean splitting and dephasing susceptibility is illustrated: the longer chain (dashed line) shows faster dephasing but
no Larmor precession. (b) Inhomogeneous dephasing time as a function of the length for disorder strength $\sigma_\mu = 2.8\text{ $\mu$eV}$ \noteandras{when disorder is uncorrelated (black) and when fully
correlated (green).} Table inset shows the \noteandras{results corresponding to uncorrelated disorder} for specific lengths.}
\label{fig:dephasingsummary}
\end{figure*}

We illustrate the dephasing dynamics by calculating 
$\braket{p_x(t)}$, the $x$ component of the
disorder-averaged polarization vector as the function of time. 
We will refer to this function as the \emph{dephasing
curve}.
We evaluate the dephasing curve based on the 
observation that the signful splitting
has a Gaussian pdf in the parameter range we
consider. 
Based on Eq.~\eqref{eq:polarization}, 
this implies the following well-known 
result\cite{MerkulovPRB2002,HansonRMP2007}
for the dephasing curve:
\begin{equation}
    \braket{p_x(t)} = e^{-\left(\frac{ \sigma_{\epsilon_0}}{\hbar} t\right)^2}\cos\left(\frac{2
    \epsilon_{0,\text{c}}}{\hbar} t\right),
\end{equation}
where $\epsilon_{0,\text{c}}$ is the clean splitting.
This result implies that dephasing follows Gaussian
decay, and this decay is 
characterized by the time scale
\begin{equation}
    \label{eq:T2star}
    T_2^* = \frac{\hbar}{\sigma_{\epsilon_0}} =\frac{\hbar}{\sigma_\mu \chi},
\end{equation}
which is often called the 
\emph{inhomogeneous dephasing time}.

Fig.~\ref{fig:dephasingsummary}a shows 
two dephasing curves for the parameter set
shown in Table \ref{tab:parameters}, the solid line showing
fast oscillations (i.e., Larmor precession),
and the dashed line showing no oscillations.
The dashed line corresponds to the diamond ($N=3083$) in Fig.~\ref{fig:splittingnumana}, 
with chain length fine-tuned such that the clean splitting 
vanishes. 
The solid line corresponds to the star ($N = 2947$)
in Fig.~\ref{fig:splittingnumana}, 
with chain length fine-tuned such that the clean
splitting has a local maximum. 
For both chain lengths, the pdf
of the signful splitting is Gaussian.
However, the mean of the signful splitting 
(which is the same as the clean splitting 
$\epsilon_{0,\text{c}}$)
is zero for the $N=3083$ case,
and finite for the $N=2947$ case, 
the latter being responsible for the oscillations
in Fig.~\ref{fig:dephasingsummary}a.
This figure also illustrates the out-of-phase relation between the clean splitting and
dephasing susceptibility (see, e.g., Fig.~\ref{fig:splittingcleananddirty}b): the 
smaller the clean splitting, the faster 
the dephasing. 

It is also interesting to note that the oscillation 
(Larmor precession)
induced by the finite clean splitting, 
as shown by the solid line in Fig.~\ref{fig:dephasingsummary}a, has
a much smaller time scale than the dephasing time. 
It would be interesting to study in detail how 
this fast Larmor precession influences the fidelity
of quantum gates, e.g., based on braiding of
MZMs\cite{AliceaNatPhys2011}.

The black solid line 
Fig.~\ref{fig:dephasingsummary}b shows the inhomogeneous dephasing time as a function of the length. 
\noteandras{(Green solid line will be discussed in the next section.)}
The dephasing time is calculated analytically by substituting the approximate formula of $\chi$ given by Eq.~\eqref{eq:approxdephasingsusc} into Eq.~\eqref{eq:T2star}. Aside from the oscillations seen 
in Fig.~\ref{fig:dephasingsummary}b, the dependence of the dephasing time on the chain length is dominated by the exponential factor
$T_2^* \propto e^{L/\xi}$ . 
The figure corresponds to a disorder strength $\sigma_\mu = 2.8\text{ $\mu$eV} $, which implies a dephasing time $T_2^* = 200\text{ ns}$ for $L/\xi=5$.
\noteandras{The oscillatory nature of the black solid result in 
Fig.~\ref{fig:dephasingsummary}b is responsible 
for the feature of Fig.~\ref{fig:dephasingsummary}a
that the shorter chain (star) has a longer $T_2^*$
than the longer chain (diamond).}

The inset of Fig.~\ref{fig:dephasingsummary}b 
shows the calculated inhomogeneous dephasing time values for specific chain lengths.
We use this table, in particular the inhomogeneous dephasing
time value $T_2^* = 200$ ns at $L/\xi =5$, to relate our
results the earlier dephasing-time estimates of  Ref.~\onlinecite{KnappPRB2018dephasing} (see Table I therein).
Ref.~\onlinecite{KnappPRB2018dephasing} predicts 
this $T_2^*$ value from intrinsic sources, without
any disorder in the sample. 
Therefore, our parameter value
$\sigma_\mu = 2.8\, \mu\text{eV}$ 
provides an estimate for the crossover
disorder strength, that is, the disorder strength above which
dephasing due to quasi-static disorder 
dominates the intrinsic dephasing mechanisms 
of a clean system
(homogeneous $1/f$ charge noise, phonons, equilibrium quasiparticles).

Experimental data indicates that the typical energy scale of 
local electrostatic fluctuations in state-of-the-art semiconductor
quantum devices is of the order of a few $\mu$eV-s, 
see, e.g., Table II of Ref.~\onlinecite{KnappPRB2018dephasing}.
This suggests that the mechanism we describe here will
be relevant for early-stage Majorana-qubit
experiments.

\section{Dephasing dynamics as a probe of spatial disorder correlations}
\label{sec:spatialdistribution}

\noteandras{
Up to this point, we have focused on the case where the on-site disorder is uncorrelated between different sites.
This model represents short-range-correlated disorder that leads to the absence of a dephasing sweet spot. 
On the other hand, if dephasing is caused by the fluctuation 
of a global control paramameter, e.g., the chemical potential, 
then a dephasing sweet spot is expected when the clean splitting
has a maximum as the function of that parameter.
This is exemplified, e.g., by Eq. (5) of Ref.~\onlinecite{KnappPRB2018dephasing}.}

\noteandras{
In this section, we go beyond the uncorrelated disorder model to
highlight the relation of the spatial correlations of the disorder and the dephasing curve. 
To this end, we generalize our disordered Kitaev-chain model by 
regarding the on-site energies as correlated normal random
variables, described by a multivariate normal distribution
with zero means and the covariance matrix
\begin{equation}
	\label{eq:covariancematrix}
	\Sigma_{ij} = \sigma_\mu^2 e^{-|i-j|a/\zeta},
\end{equation}
where $i,j$ are site indices and $\zeta$ is the correlation length. For further details of the model see Appendix \ref{appendix:correlateddisorder}, for a discussion between our model and disorder in real devices see Sec.~\ref{sec:discussion}. Parameter $\zeta$ controls the spatial correlation in the disorder realizations: $\zeta \lesssim a$ indicates uncorrelated disorder, furthermore $\zeta \gtrsim L$ corresponds to the homogenous, fully correlated case.}

\begin{figure*}
\includegraphics[width=17.8cm]{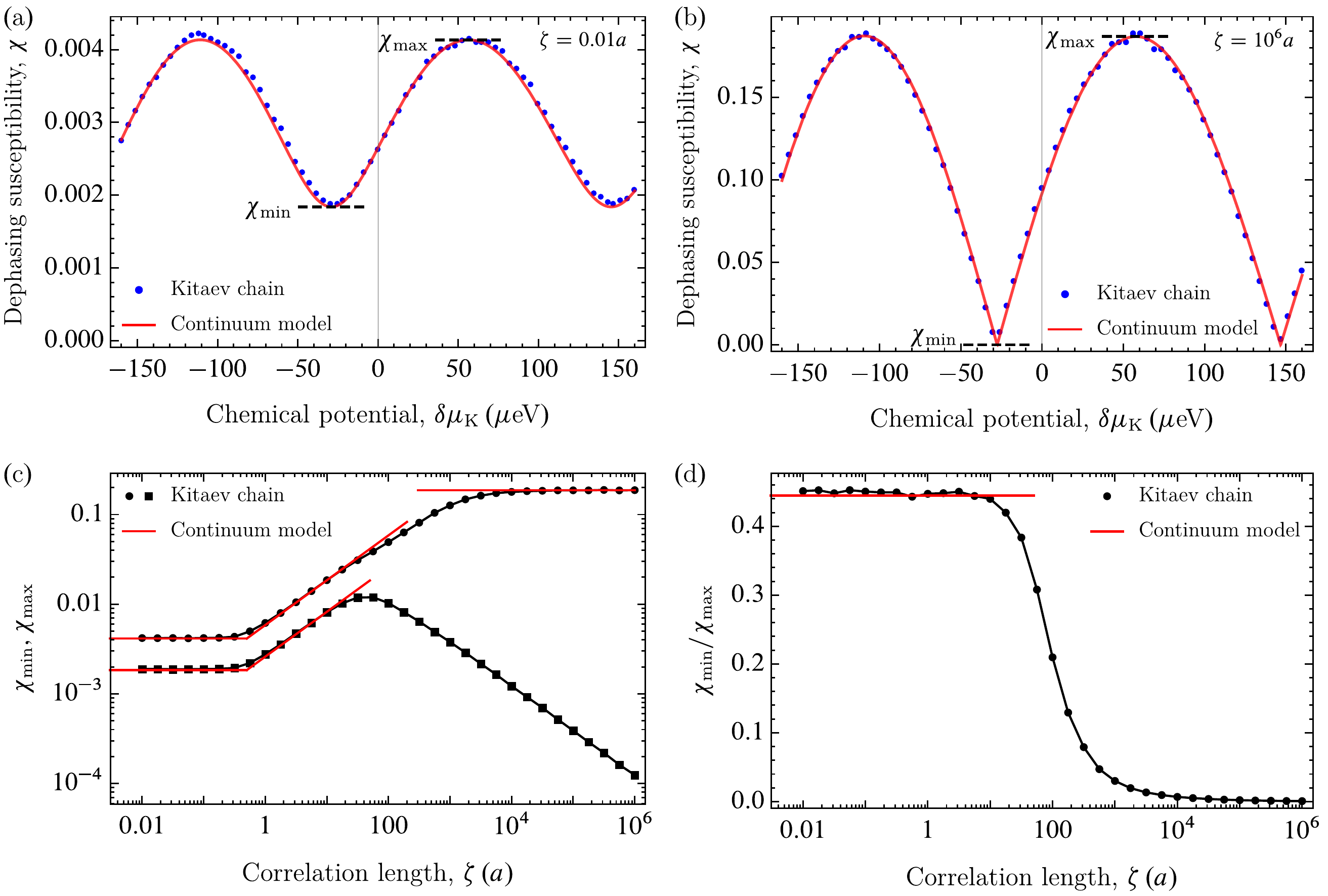}
\centering
\caption{\noteandras{Effect of the spatial correlation of disorder on dephasing. (a-b) Dephasing susceptibility as a function of the chemical potential for  uncorrelated disorder (a) and for fully correlated disorder (b). Points show results from the Kitaev chain model with spatially correlated disorder. Solid lines show analytical results of the continuum model in the two limits. (c) Adjacent minimum and maximum values of the dephasing susceptibility as a function of the correlation length. (d) Ratio of the minimum and the maximum values as a function of the correlation length. This ratio is constant in the uncorrelated and weakly correlated regimes ($\zeta \lesssim 10a$), and goes to zero as the correlation length increases. The latter feature corresponds to the dephasing sweet spots of the fully correlated regime seen in (b).}
}
\label{fig:spatialcorrelation}
\end{figure*}

\noteandras{Fig.~\ref{fig:spatialcorrelation}a and b show the dephasing suscesctibility as a function of the chemical potential for $\zeta=0.01a$ (uncorrelated disorder) and $\zeta=10^6a$ (fully correlated disorder), respectively. System size is fixed to $3000$ sites. The chemical potential $\delta\mu_\text{K}$  is measured from $\mu_\text{K}$ given in Tab.~\ref{tab:parameters}. Blue points correspond to a numerical calculation based on the Kitaev chain model with correlated on-site energy disorder. Red line of Fig.~\ref{fig:spatialcorrelation}a shows the analytically obtained dephasing  susceptibility of the continuum model, see Eq.~\eqref{eq:approxdephasingsusc}, with $a_\text{dis}=a$. Red line in Fig.~\ref{fig:spatialcorrelation}b shows the dephasing susceptibility of the continuum model against homogenous chemical potential disorder, that can be obtained by taking the derivative of the clean splitting formula in Eq.~(\ref{eq:pientkageneralized}a) with respect to $\mu_\text{K}$. We find furthermore that the corresponding lengthy formula can be approximated (not shown) as
\begin{equation}
	\label{eq:chifullycorr}
	\chi_\text{fcorr} = \frac{2L}{\xi}\frac{k_\text{F}^2}{k_\text{F}^2-1/\xi^2} e^{-L/\xi}\left|\cos\left(\sqrt{k_\text{F}^2-1/\xi^2}L\right)\right|.
\end{equation}
}

\noteandras{
The main observations in Fig.~\ref{fig:spatialcorrelation}a and b
are as follows: (i) the dephasing susceptibility oscillates as a function of the chemical potential in both panels, (ii) the magnitude of the oscillations is greater in the case of fully correlated disorder (Fig.~\ref{fig:spatialcorrelation}b) than in the case of uncorrelated 
disorder (Fig.~\ref{fig:spatialcorrelation}a), 
(iii) the case of fully correlated disorder (Fig.~\ref{fig:spatialcorrelation}b)
exhibits dephasing sweet spots, where the dephasing susceptibility
$\chi$ vanishes.
See, e.g., at $\delta \mu_\text{K} \approx -30\, \mu\text{eV}$.}

\noteandras{In a future Majorana-qubit dephasing experiment, 
oscillations such as those shown in Fig.~\ref{fig:spatialcorrelation}a and b
are directly observable, e.g., by tuning the chemical potential 
via a back-gate voltage. 
Here we argue that such an oscillatory data set can be used 
to infer the disorder correlation length.}

\noteandras{
To illustrate this opportunity, we take two adjacent extrema
in Fig.~\ref{fig:spatialcorrelation}a and b: 
a minimum ($\chi_\text{min}$) and a maximum ($\chi_\text{max}$), 
located closest to $\delta \mu_\text{K} = 0$ (see labels in figure).
In Fig.~\ref{fig:spatialcorrelation}c, we show how $\chi_\text{min}$ and $\chi_\text{max}$ evolve as functions of the correlation length $\zeta$. Results from the Kitaev chain model are depicted by black markers, 
whereas red lines show the analytical results in the uncorrelated regime ($\zeta \lesssim a$), the weakly correlated regime ($a \lesssim \zeta \lesssim 10a$, derivation discussed below), and the fully correlated regime ($\zeta \gtrsim L$). For uncorrelated disorder ($\zeta \lesssim a$), the extremal dephasing susceptibilies are approximately constants. However, for weakly correlated disorder ($a \lesssim \zeta \lesssim 10a$), $\chi_\text{min}$ and $\chi_\text{max}$ increase as the correlation length is increased. This  means that qubit dephasing is more sensitive to correlated disorder than to uncorrelated one. In the fully correlated disorder limit ($\zeta \gtrsim L$), $\chi_\text{max}$ saturates, whereas $\chi_\text{min}$ decreases, in accordance with the dephasing sweet spot seen for this limit in Fig.~\ref{fig:spatialcorrelation}b.} 

\noteandras{To introduce a procedure which estimates the correlation length from the dephasing curves, we plot the ratio $\chi_\text{min}/\chi_\text{max}$ as a function of the correlation length in Fig.~\ref{fig:spatialcorrelation}d. In the uncorrelated and weakly correlated regimes ($\zeta \lesssim 10a$), the ratio is a constant, for longer correlation length, it tends to zero due to the existence of dephasing sweet spots.}

\noteandras{The ratio $\chi_\text{min}/\chi_\text{max}$ as a function of the correlation length is monotonic, which provides an opportunity to characterize the correlation length experimentally, in the following way: one can measure the dephasing curves by varying the chemical potential and determine the corresponding dephasing times. By fine tuning the chemical potential, two adjacent minima and maxima of the dephasing times can be determined: $T_{2,\text{min}}^*$ and $T_{2,\text{max}}^*$. The ratio of the extremal dephasing times equals to the inverse ratio of the extremal dephasing susceptibilities, i.e., $T_{2,\text{max}}^*/T_{2,\text{min}}^*=\chi_\text{min}/\chi_\text{max}$, which can be seen from Eq.~\eqref{eq:T2star}. Using Fig.~\ref{fig:spatialcorrelation}d, one can infer the correlation length, or at least can distinguish between short-range and long-range disorder correlations.}

\noteandras{In order to support our numerical results in the uncorrelated, weakly, and fully correlated regimes in Fig.~\ref{fig:spatialcorrelation}c, we provide analytical results from the continuum model, shown as the three solid red line segments in Fig.~\ref{fig:spatialcorrelation}c. In the uncorrelated limit ($\zeta \lesssim a$), we make use of the extrema of Eq.~\eqref{eq:approxdephasingsusc} with $a_\text{dis}=a$ in order to determine $\chi_\text{min}$ and $\chi_\text{max}$. In the fully correlated regime ($\zeta \gtrsim L$), we take the maximum of Eq.~\eqref{eq:chifullycorr}.}

\noteandras{To obtain an analytical result for the weakly correlated regime ($a \lesssim \zeta \lesssim 10a$) from the continuum model, we make use of Eq.~\eqref{eq:approxdephasingsusc}, which expresses the dephasing susceptibility $\chi$ as function of the parameter $a_\text{dis}$. (Recall that in our continuum model, the disorder is modeled by series of potential steps of length $a_\text{dis}$, see Eq.~\eqref{eq:contdisorder}). We substitute $a_\text{dis} = 2 \zeta$ in Eq.~\eqref{eq:approxdephasingsusc} to express $\chi$ as the function of the disorder correlation length $\zeta$. In what follows, we argue why we identify $a_\text{dis}$ with $2 \zeta$.} 

\noteandras{The covariance function of the disorder in the 
continuum model can be written as
\begin{equation}
	C_\text{C}(x,y) = \begin{cases}
				\sigma_\mu^2, & \text{if $\left\lfloor x/a_\text{dis} \right\rfloor = \left\lfloor y/a_\text{dis} \right\rfloor$},\\
            	0, & \text{otherwise}.
		     \end{cases}
\end{equation} 
Matching is based on the following relation
\begin{equation}
	\label{eq:matchcorrelationscales}
	\frac{\int_0^\infty{x\,C_\text{C}(x,0)}\text{d}x}{\int_0^\infty{C_\text{C}(x,0)}\text{d}x} =
	\frac{\int_0^\infty{x\,C_\text{K}(x,0)}\text{d}x}{\int_0^\infty{C_\text{K}(x,0)}\text{d}x},
\end{equation}
where we use the continuum form of the covariance matrix $\Sigma_{ij}$ [cf.~Eq.~\eqref{eq:covariancematrix}]:
\begin{equation}
	C_\text{K}(x,y) = \sigma_\mu^2 e^{-|x-y|/\zeta}.
\end{equation}
Eq.~\eqref{eq:matchcorrelationscales} leads to $a_\text{dis}=2\zeta$. By substituting it into Eq.~\eqref{eq:approxdephasingsusc}, we find good agreement between the numerical (black points) and analytical (red solid lines) results, see the weakly correlated disorder regime ($a \lesssim \zeta \lesssim 10a$) in Fig.~\ref{fig:spatialcorrelation}c.}

\noteandras{Finally, we discuss the dephasing time for fully correlated disorder, that is shown in Fig.~\ref{fig:dephasingsummary}b by the green solid line, as a function of the length. The dephasing time is calculated analytically by substituting Eq.~\eqref{eq:chifullycorr} into Eq.~\eqref{eq:T2star}. Results from fully correlated disorder oscillates in phase with the results arising from uncorrelated disorder (black solid line). Dephasing sweet spots appear as singularities, showing diverging dephasing time. This is the consequence of our limited dephasing model which is based on the linear approximation $1/T_2^* \propto \sigma_\mu$, see Eq.~\eqref{eq:T2star}. A higher-order approach would resolve the singular behaviour.}

\noteandras{In conclusion, our results in Fig.~\ref{fig:dephasingsummary} provide important practical insights on how to optimize a Majorana qubit setup for a dephasing experiment. The effect of homogeneous charge noise, that is, a uniform random shift of the chemical potential, can be mitigated by fine-tuning the chemical potential: the dephasing time can be strongly enhanced by such a fine tuning. If the noise is not spatially homogeneous, then the magnitude of the improvement depends on the correlation length of the disorder: for uncorrelated noise, fine tuning could yield at most a factor of two improvement (cf.~Fig.~\ref{fig:spatialcorrelation}a), but this improvement factor gradually increases for increasing disorder correlation length.}

\section{Discussion}
\label{sec:discussion}

In the main part of this work, we used a model of 
short-range-correlated disorder. 
This is admittedly a minimal model of disorder in real samples, 
nevertheless we find it important and relevant to provide the 
corresponding results, because 
(i) this is a conceptually simple, 
canonical model, often used in the literature,
applied to effects ranging from Anderson localisation to
Majorana physics\cite{BrouwerPRL2011,HegdePRB2016}, and
(ii) these results also serve as benchmark for more realistic models.
Also, the level of disorder
in state-of-the-art hybrid nanowires seems to be too strong
to allow for the clear observation of Majorana zero modes,
which suggest that disorder will likely play a dominant role
also in the initial Majorana-qubit experiments, 
e.g., qubit dephasing time measurements.
In real nanowire samples, disorder might arise due to various
physical mechanisms\cite{GiustinoJOPMat2021}, e.g., fluctuating 
charge traps in the substrate, atoms, ions, molecules contaminating
the wire surface, impurity atoms built in to the crystal upon growth,
electron scattering on rough or oxidized wire surface and 
core-shell interface, inhomogeneous strain patters due
to thermal expansion coefficient mismatch and
metal deposition (shell, gates, contacts), gate-voltage
fluctuations, etc.
It is an important ongoing effort to mitigate these mechanisms;
alternatively, it is useful to characterize and control their effects 
on Majorana qubit decoherence.

In our dephasing calculation, 
we have chosen the quasistatic approximation also for 
its conceptual simplicity and widespread use in the literature\cite{TosiNatComm2017,BoterPRB2020}.
In real devices, classical or quantum noise
often follows a characteristic noise spectrum,
e.g., $1/f$ noise\cite{DialPRL2013,FreemanAPL2016,YonedaNatNano2018,Makhlin,KnappPRB2018dephasing,HuangPRA2019,HetenyiPRB2019,
MishmashPRB2020,CywinskiPRB2020,KhindanovScipost2021}, Johnson-Nyquist noise, 
quantum noise of phonons\cite{KnappPRB2018dephasing,AseevPRB2018}, 
gate-voltage fluctuations\cite{SchmidtPRB2012,AseevPRB2019,KnappPRB2018dephasing,MishmashPRB2020}, etc. 
Going beyond the quasistatic approximation
by incorporating these frequency-dependent noise features
would be an important addition to this work. 
An especially appealing task is to describe
the combined effect of static spatial disorder and 
fluctuating electric fields; this direction might
actually reveal connections between actual device
physics and the minimal model used in our present work. 
A conceptually different, but equally important information loss
mechanism for Majorana qubits is 
quasiparticle poisoning\cite{BudichPRB2012,RainisPRB2012,ColbertPRB2014,KarzigPRB2017,AlbrechtPRL2017,KarzigPRL2021}.

In this work, we focused on the case of low disorder, in the hope
that material growth and device fabrication advances will convey
qubit experiments in that parameter range. 
Current devices might have much stronger disorder\cite{PikulinNJP2012,HainingPanPRR2020,woodsarxiv2021} and it is 
an interesting extension of our work to study how Majorana qubit
dephasing occurs in the presence of strong disorder.
A further natural extension of our work is to step-by-step move
from the Kitaev-chain minimal model to more realistic
real-space models, e.g., from 1D Rashba wire\cite{OregPRL2010,LutchynPRL2010,panPRB2021}
to 3D Schrödinger-Poisson 
models\cite{VuikNJP2016,AntipovPRX2018,woodsarxiv2021}, and beyond.

\noteandras{
One of the key result of the paper is that the
dephasing susceptibility oscillates
as the function of system parameters
out-of-phase with respect to the oscillations of
the clean splitting.
This is shown in Fig.~\ref{fig:splittingcleananddirty}b.
How robust is this result upon relaxing the simple hard-wall
boundary condition leading to the result in Fig.~\ref{fig:splittingcleananddirty}b?
We have performed numerical simulations exploring 
this question, by extending our model in two ways: (1) We have relaxed the hard-wall boundary condition to a confinement potential that has a step-like dependence at the two edges of the 1D topological superconductor, (2) we have added a homogeneous electric field, that is, a chemical potential that varies linearly with position. In the parameter range we studied, the two quantities were following the same type of out-of-phase oscillations as shown in Fig.~\ref{fig:splittingcleananddirty}b. 
We see it as an interesting follow-up question to understand this robustness. }

\section{Conclusions}
\label{sec:conclusions}

We have studied the Majorana splitting of the
disordered 
topological Kitaev chain, serving as a minimal 
model of dephasing of Majorana qubits. 
\noteandras{Focusing on the case of spatially uncorrelated disorder, we} characterized the Gaussian
probability distributions of the signful
splitting, using numerics as well as
simple semi-analytical and approximate analytical
techniques. 
We established a Gaussian decay envelope
for the dephasing curve, as a consequence
of the Gaussian distribution of the signful 
splitting. 
We have found that the standard deviation of the signful
splitting, and hence the dephasing rate, oscillates
as the function of system parameters
out-of-phase with respect to the oscillations of
the clean splitting. 
We have also pointed out the
absence of dephasing sweet spots \noteandras{in the case of 
spatially uncorrelated quasistic disorder.}
\noteandras{Furthermore, we have described how Majorana qubit dephasing changes as the function of disorder correlation length, and argued that  dephasing measurements can be used to characterize the disorder correlation length.}
We expect that our results will 
be used in the design
and interpretation of future 
experiments, aiming to demonstrate topologically 
protected quantum memory, quantum
dynamics, or quantum computing, based
on Majorana zero modes.

\acknowledgments

We thank J.~Asb\'oth, P.~Brouwer,
L.~Oroszl\'any, A.~Romito, and G.~Sz\'echenyi
for helpful discussions, and G.~Takács for computational resources.
This research was supported by the Ministry of Innovation and Technology and the National Research, Development and Innovation Office (NKFIH) within the Quantum Information National Laboratory of Hungary,
the BME Nanotechnology and 
Materials Science TKP2020 IE grant (BME IE-NAT TKP2020),
the Quantum Technology National Excellence Program (Project No. 2017-1.2.1-NKP-2017-00001),
and the OTKA Grants FK124723 and FK132146.

\appendix

\section{Connecting the continuum model with the Kitaev chain}
\label{appendix:connecttwomodel}

\begin{figure*}
    \includegraphics[width=17.8cm]{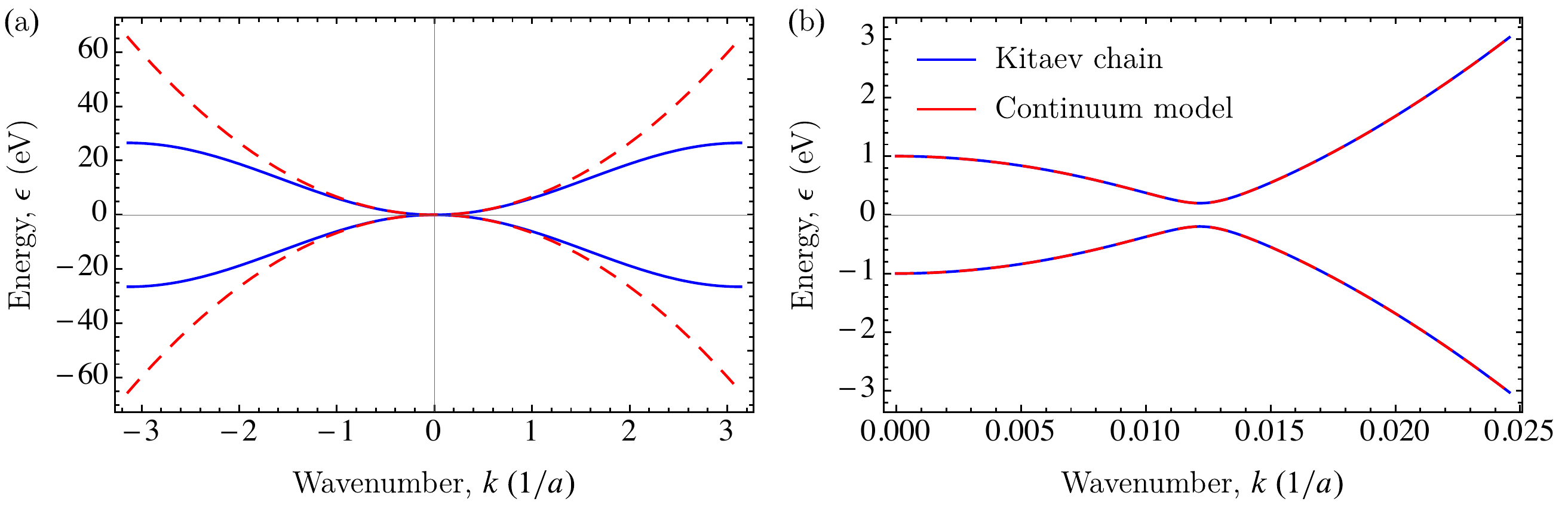}
    \centering
    \caption{The Kitaev chain band structure (blue) 
    and the continuum model band structure (red),
    (a) over the 1D Brillouin zone,
    (b) in the vicinity of zero energy,
    with parameters 
    chosen such that the band structures match each other in the vicinity of zero energy.
    See Table \ref{tab:parameters} for
    parameter values.
    \label{fig:Kitaev_vs_cont}}
\end{figure*}

The Kitaev chain [Eq.~\eqref{eq:kitaevchain}] is a discretized version
of the continuum model [Eq.~\eqref{eq:contdisorder}],
and vice versa, the continuum model can be obtained from the 
Kitaev chain via the envelope-function approximation.
The relation of the two models is outlined in Ref.~\onlinecite{PientkaNJP2013}, but for the sake of self-containedness, we describe it here in detail. We match the parameters of the two models
via matching their dispersion relations as shown in Fig.~\ref{fig:Kitaev_vs_cont}.

To match the two models, we recall the BdG Hamiltonian of the Kitaev chain in momentum 
space, which reads
\begin{equation}
    \label{eq:Kitaevmomentumspace}
    \mathcal{H}_\text{K}(k) = (-2t\cos(k a) -\mu_\text{K})\sigma_z + 2\Delta_\text{K}\sin(k a) \sigma_y.
\end{equation}

Here we note that the momentum-space superconducting term 
$2\Delta_\text{K} \sin(ka) \sigma_y$ of the Kitaev chain
is proportional to $\sigma_y$, 
whereas the corresponding term $\Delta'_\text{C} \hbar k \sigma_x$ of
the continuum model [Eq.~\eqref{eq:continuum}] is proportional to $\sigma_x$.
This difference is irrelevant, can be
transformed away with a unitary 
transformation in Nambu space,
since the rest of both Hamiltonians
is proportional to $\sigma_z$.

Matching the continuum model and the Kitaev chain model 
is based on the following criteria:

\begin{subequations}
\begin{enumerate}
    \item 
    The lengths in the two models are naturally
    matched as $L = N a$, where $L$ is the 
    length of the wire and $N$ is the number
    of sites in the lattice model.
	\item In the absence of superconductivity, 
	the effective mass in the vicinity of $k=0$ has to be the same 
	in the two models, 
    which yields the condition:
	\begin{equation}
		m = \frac{\hbar^2}{2t a^2}.
		\label{match1}
	\end{equation}
	
	\item In the absence of the superconducting terms, the minima
	of the bulk spectra have to be at the same energy.
	This is achieved by adjusting the chemical potentials
	in the following way:
	\begin{equation}
		\mu_\text{C} = 2t + \mu_\text{K}.
		\label{match2}
	\end{equation}
	
	\item
	The low-energy (close to zero energy) spectra
	of the two models will be similar if 
	the minimum of the bulk normal band 
	is just slightly below zero energy;
	formally this can be written as
	\begin{equation}
	    0 < 1+ \frac{\mu_\text{K}}{2t} \ll 1.
	    \label{match3}
	\end{equation}
		
	\item In the presence of superconductivity, the
	superconducting gaps have to be equal, 
	a condition approximately satisfied by the identification
	\begin{equation}
		\Delta_\text{C}' = \frac{\Delta_\text{K}a}{\hbar}\sqrt{2-\frac{\mu_\text{K}}{t}}.
		\label{match4}
	\end{equation}
	We note that here we have already assumed that Eqs.~\eqref{match1} and \eqref{match2} are fulfilled. Eq.~\eqref{match4} is an approximation in the sense that we match energy gaps of the two models that are opened at $k_\text{F}$, 
	i.e., at the wave number where the band touches zero in the absence of the superconductivity. The actual gap (i.e., the energy difference
	minimized over the wave number)
	is in general located at a slightly different
	wave number $k_0$, but in the limit of Eq.~\eqref{match3}, $k_0\approx k_\text{F}$.
	\end{enumerate}
\end{subequations}
Based on the above criteria, 
we choose the parameter values 
listed in Table \ref{tab:parameters}
to compare the results of the Kitaev chain and the continuum model.

The energy dispersion of the Kitaev
chain (blue solid) and that of the 
continuum model (red dashed)
are compared over the 1D Brillouin zone
in Fig.~\ref{fig:Kitaev_vs_cont}a,
and in the vicinity of the Brillouin 
zone center and the Fermi wave number
in Fig.~\ref{fig:Kitaev_vs_cont}b.

Below, we will need the following relations 
between the parameters of the two models:
\begin{subequations}
\begin{align}
    \label{eq:k_F_Kitaev}
    k_\text{F} &= \frac{\sqrt{2+\frac{\mu_\text{K}}{t}}}{a},\\
    \label{eq:Delta_C_Kitaev}
    \Delta_\text{C} &=
    \Delta_\text{K}\sqrt{4-\left(\frac{\mu_\text{K}}{t}\right)^2}, \\
    \label{eq:xi_Kitaev}
    \xi &= \frac{2 t a}{\Delta_\text{K}\sqrt{2-\frac{\mu_\text{K}}{t}}}.
\end{align}
\end{subequations}
We obtain Eq.~\eqref{eq:k_F_Kitaev} from Eq.~\eqref{eq:k_F} by substituting Eqs.~\eqref{match1} and \eqref{match2}. We get Eq.~\eqref{eq:Delta_C_Kitaev} from Eq.~\eqref{eq:Delta_C} by substituting Eqs.~\eqref{match4} and \eqref{eq:k_F_Kitaev}. 
We obtain Eq.~\eqref{eq:xi_Kitaev} from Eq.~\eqref{eq:xi} by combining Eqs.~\eqref{eq:v_F}, \eqref{match1}, \eqref{eq:k_F} and \eqref{eq:Delta_C}.

\section{Inferring the standard deviation of the signful splitting from samples of the splitting}
\label{appendix:signfulstd}

Figure \ref{fig:splittingcleananddirty}b shows the standard deviation of 
$\sigma_{\epsilon_0}$ of
the signful splitting $\epsilon_0$ of a Kitaev chain
due to disorder.
How did we compute $\sigma_{\epsilon_0}$?
The smallest non-negative eigenvalue of the BdG matrix is
the \emph{absolute value} of the signful splitting, hence its standard deviation taken over many disorder realizations does not provide
$\sigma_{\epsilon_0}$.
Here, we provide an indirect way to compute $\sigma_{\epsilon_0}$
by assuming that the signful splitting is normally distributed, 
an assumption in accordance with our result \eqref{eq:gaussian}. 
Under that assumption, the absolute value of the signful splitting has \emph{folded normal distribution}. 
We have an easy access to samples of the splitting by using BdG Hamiltonian, and by following the procedure outlined below, 
we are able to compute $\sigma_{\epsilon_0}$
from samples of the splitting. 

Let us use a general notation for easier readability. 
The probability density function of the normal random variable $X$, 
representing the signful splitting, 
reads
\begin{equation}
    f_{X}(x) = \frac{1}{\sqrt{2\pi}\sigma}e^{-\frac{\left(x-m\right)^2}{2\sigma^2}},
\end{equation}
where $m$ is the mean and $\sigma$ is the standard deviation of $X$. 
The probability density function of the random variable 
$|X|$, representing the splitting, is
\begin{equation}
    f_{|X|}(x) = \frac{1}{\sqrt{2\pi}\sigma}e^{-\frac{\left(x-m\right)^2}{2\sigma^2}}\left(1+e^{\frac{2 m x}{\sigma^2}}\right),
\end{equation}
which is often called a folded normal distribution.

We estimate the parameters $m$ and $\sigma$ 
(mean and standard deviation of signful splitting)
from a sample $\{x_i | i = 1, \dots n\}$  
of $|X|$ (the splitting). 
Here $n$ is the size of the sample. 
Our estimation 
is based on the maximum likelihood estimation procedure. 
The log-likelihood of the distribution
estimated from the sample $\{x_i\}$ 
can be written as
\begin{align}
    l(\{x_i\};m,\sigma) &= \log\left[\prod_{i=1}^{n}{f_{|X|}(x_i)}\right]= \nonumber \\
    & -\frac{n}{2}\log\left(2\pi\sigma^2\right) - \sum_{i=1}^{n}{\frac{\left(x_i-m\right)^2}{2\sigma^2}} \nonumber \\
    &+\sum_{i=1}^{N}{\log\left(1+e^{\frac{2 m x_i}{\sigma^2}}\right)}.
\end{align}

To estimate the value of $m$ and $\sigma$, we need to find
the maximum point of the likelihood function, hence we take $\partial_m{l(\{x_i\};m,\sigma)}=0$ and $\partial_\sigma{l(\{x_i\};m,\sigma)}=0$, that lead to
\begin{subequations}
    \begin{align}
        m &= \frac{1}{n}\sum_{i=1}^{n}{x_i\tanh{\left(\frac{m x_i}{\sigma}\right)}}, \label{eq:appendix_mean}\\
        \sigma^2 &= m^2 + \frac{1}{n}\sum_{i=1}^{n}{x_i^2} - \frac{2m}{n}\sum_{i=1}^{n}{x_i\tanh{\left(\frac{m x_i}{\sigma}\right)}}.
        \label{eq:appendix_std}
    \end{align}
\end{subequations}
From Eqs.~\eqref{eq:appendix_mean} and \eqref{eq:appendix_std}, we get
\begin{equation}
    \sigma^2 = \left(\frac{1}{n}\sum_{i=1}^{n}{x_i^2}\right) - m^2.
    \label{eq:appendix_std_2}
\end{equation}

In general, the coupled Eqs.~\eqref{eq:appendix_mean} and \eqref{eq:appendix_std} have to be solved. 
However, in our case, to determine the standard deviation of the signful splitting, Eq.~\eqref{eq:appendix_std_2}, is sufficient as 
know the square of the signful splitting mean $m$:
it is equal to the square of the splitting of the clean system
$\varepsilon_{0,\text{c}}$. This implies the formula
\begin{equation}
    \sigma_{\epsilon_0} = \sqrt{\left(\frac{1}{n}\sum_{i=1}^{n}{\varepsilon_{0,i}^2}\right) - \varepsilon_{0,\text{c}}^2},
    \label{eq:appendix_splitting_std}
\end{equation}
where $\varepsilon_{0,i}$-s are splittings in disordered realizations and $\varepsilon_{0,\text{c}}$ is the splitting for the clean system. 
We used this result to compute the data in 
Fig.~\ref{fig:splittingcleananddirty}b.

\section{Comparison with the results of Brouwer et al. PRL 2011}
\label{appendix:compareBrouwer}

In the main text, we predict a normal distribution for signful splitting $\epsilon_0$.
On the other hand, the key result of Ref.~\onlinecite{BrouwerPRL2011} is that the
\emph{splitting envelope} $\varepsilon_{0,\text{max}}$ (for clarification, see their Fig.~1c)
has a log-normal distribution. 
Although the two quantities (signful splitting and splitting envelope) are
not the same, they are in fact interrelated.
In this appendix, we identify a parameter range where
both our results and the results of Ref.~\onlinecite{BrouwerPRL2011}
are valid, and establish the relation of these results.
Our comparison suggests that the two unknown constants
appearing in the analytical results of Ref.~\onlinecite{BrouwerPRL2011} 
($C_\text{m}$ and $C_\text{v}$, see below)
are actually zero.

The
main result of Ref.~\onlinecite{BrouwerPRL2011} is as follows. 
The quantity
$\ln(\varepsilon_{0,\text{max}}/2\Delta_\text{C})$ has a normal distribution with mean and variance given by
their Eq.~(16), that is, 
\begin{subequations}
\label{eq:brouwer}
\begin{align}
	\braket{\ln(\varepsilon_{0,\text{max}}/2\Delta_\text{C})} &= -L\left[1/\xi- 1/2l\right] + C_\text{m}, \\
	\text{var}\ln(\varepsilon_{0,\text{max}}/2\Delta_\text{C}) &= L/2l + C_\text{v}.
\end{align}
\end{subequations}
Here, $C_\text{m}$ and $C_\text{v}$ are the unknown constants, 
that is, unknown order-of-unity corrections independent of $L$, $l$ and $\xi$.
\noteandras{(Even though these constants are not displayed in Eq. (16) of 
Ref.~\onlinecite{BrouwerPRL2011}, they are introduced in the text
following that equation.)}
Furthermore, $l = \hbar^2 v_\text{F}^2 / \gamma$ is the mean free path, where $\gamma$ corresponds to the disorder strength in their model, which is identified with our model as $\gamma = a_\text{dis} \sigma_\mu^2$.

Their results stand if the following conditions are satisfied:
\begin{subequations}
    \label{eq:rangeofvalidity}
    \begin{align}
        1/k_\text{F} & \ll \xi, \\
        \xi & < 2l, \\
        \varepsilon_{0,\text{max}} &\ll \text{min}(\Delta_\text{C},\hbar/\tau),
    \end{align}
\end{subequations}
where $\tau=\hbar v_\text{F}/l$. 
On the other hand, our result for the clean splitting
\eqref{eq:pientkageneralized} is valid 
if $L \gg \xi$, 
and our result for the dephasing susceptibility 
to disorder \eqref{eq:approxdephasingsusc} is 
valid if $L \gg \xi$ and if disorder is weak.

First, we assume that the parameter range of validity 
of the two results have some overlap, and 
show that in such a common parameter range, 
the two results are consistent. 
Second, we provide an example for the common 
parameter range where both results should be valid
and hence should be consistent with each other. 

To show
the consistency of the two results, we suppose that
\begin{subequations}
    \label{eq:lognormalnormalassumptions}
    \begin{align}
        l &\gg \xi, \\
        L/2l &\ll 1, \\
        C_\text{m} &= 0, \\
        C_\text{v} &= 0.
    \end{align}
\end{subequations}
Eq.~(\ref{eq:lognormalnormalassumptions}a) stands for weak disorder, 
whereas Eq.~(\ref{eq:lognormalnormalassumptions}b) 
together with Eq.~(\ref{eq:lognormalnormalassumptions}d) provides that $\ln(\varepsilon_{0,\text{max}}/2\Delta_\text{C})$ has a standard deviation much
smaller than one.
Furthermore, the choice of $C_\text{m}$ and $C_\text{v}$ in Eqs.~(\ref{eq:lognormalnormalassumptions}c-d) is required to match the result of Ref.~\onlinecite{BrouwerPRL2011} with our results.

Our results, 
together with Eqs.~\eqref{eq:lognormalnormalassumptions} imply that
the splitting envelope 
$\varepsilon_{0,\text{max}}$ approximately follows normal distribution with mean and standard deviation as follows:
\begin{subequations}
\label{eq:Brouwersimp}
\begin{align}
	\braket{\varepsilon_{0,\text{max}}} &= 2\Delta_\text{C}e^{-L/\xi} , \\
	\sigma_{\varepsilon_{0,\text{max}}} &= \Delta_\text{C}\sqrt{\frac{2L}{l}}e^{-L/\xi}=\sigma_\mu \frac{\sqrt{2La_\text{dis}}}{\xi}e^{-L/\xi}.
\end{align}
\end{subequations}
We obtained Eq.~(\ref{eq:Brouwersimp}a), 
from Eq.~(\ref{eq:pientkageneralized}a)
by taking the  limit $k_\text{F} \ll 1/\xi$ and by omitting the sinusoidal oscillatory part in the latter.
Furthermore, we obtained Eq.~(\ref{eq:Brouwersimp}b) 
from Eq.~\eqref{eq:approxdephasingsusc}
by taking the limit $L\gg\xi$, and by substituting the cosine term with $-1$. 
The latter substitution is needed because
the disorder-induced 
standard deviation of the splitting has a local minimum
whenever the clean splitting has a local maximum (see Fig.~\ref{fig:splittingnumana}a).

The key mathematical 
statement we use to show the consistency of
Eq.~\eqref{eq:brouwer}
and
Eq.~\eqref{eq:Brouwersimp}
is the
following: 
If $X$ is a log-normal random variable 
such that 
$\ln X$ is a normal random variable with
mean $\mu$ and standard deviation $\sigma$
(that is, $\ln X\sim\mathcal{N}(\mu,\sigma)$),
and the standard deviation fulfills $\sigma \ll 1$, 
then 
$X$ is approximately a normal random variable 
with mean $e^\mu$ and standard deviation
$e^\mu \sigma$ (that is, 
$X \sim \mathcal{N}(e^\mu,e^\mu \sigma)$).
This follows from the fact that the exponential 
function can be well approximated around any
point by its linear series expansion in a sufficiently 
small environment of the point.
We apply this approximation to 
Eq.~\eqref{eq:brouwer} using
the assumptions of Eq.~\eqref{eq:lognormalnormalassumptions}.
This procedure yields
Eq.~\eqref{eq:Brouwersimp},
implying that our result is consistent 
with the earlier result. 

Finally, we provide an example for the common
parameter range where both results are valid.
Equation (\ref{eq:rangeofvalidity}a) is satisfied for the parameter set in Table \ref{tab:parameters}. In the weak disorder limit, Eq.~(\ref{eq:rangeofvalidity}b) is fulfilled. For weak disorder $\hbar/\tau\ll\Delta_\text{C}$, furthermore using Eq.~(\ref{eq:Brouwersimp}a), 
the condition 
$\varepsilon_{0,\text{max}} \ll \hbar/\tau$ is equivalent to the condition
\begin{equation}
    \label{eq:lenghtvalidity}
    \ln(2l/\xi) \ll L/\xi.
\end{equation}
In addition, Eq.~\eqref{eq:lenghtvalidity} and Eq.~(\ref{eq:lognormalnormalassumptions}d) can be combined as
\begin{equation}
    \label{eq:lenghtvalidityfull}
    \ln(2l/\xi) \ll L/\xi \ll 2l/\xi.
\end{equation}
For weak disorder, there is a finite interval
for the system length $L$ 
where Eq.~\eqref{eq:lenghtvalidityfull} is fulfilled. 
For example, for
the parameter values given in Table~\ref{tab:parameters}, and for disorder strength $\sigma_\mu = 10\,\mu\text{eV}$, Eq.~\eqref{eq:lenghtvalidityfull} is
evaluated
\begin{equation}
    14500 \ll L/a \ll 4.32 \times 10^{10}.
\end{equation}
Note that our numerical results shown in the
main text correspond to system lengths
that are one order of magnitude smaller 
than the lower end of this interval. 

To conclude, we have established the consistency 
between the earlier analytical results of
Ref.~\onlinecite{BrouwerPRL2011} for the 
statistics of the splitting envelope, and
our analytical results for the
statistics of the signful splitting 
described in the main text.
To ensure this consistency, 
we had to assume that the order-of-unity
constant offset parameters $C_\text{m}$
and $C_\text{v}$, which were not calculated
in Ref.~\onlinecite{BrouwerPRL2011}, 
are actually zero.
This indirect determination 
of the offset parameters 
is a useful byproduct of the comparison.

\section{Correlated disorder}
\label{appendix:correlateddisorder}

\noteandras{In Sec.~\ref{sec:spatialdistribution}, we study the effect of the spatial correlations of the disorder on dephasing. To determine the dephasing susceptibility of the disordered Kitaev chain, we have to generate numerous spatially correlated disorder realizations. In this appendix, we show a method allowing us to do that in an efficient way.}

\noteandras{Vector of the on-site energies $\boldsymbol{\delta\mu}_\text{K}$ is an $N$-dimensional random variable vector described by a multivariate normal distribution, i.e., $\boldsymbol{\delta\mu}_\text{K}\sim\mathcal{N}(\mathbf{0},\mathbf{\Sigma})$, where 
\begin{equation}
	\label{eq:covmatrixappendix}
	[\mathbf{\Sigma}]_{ij} = \sigma_\mu^2 e^{-|i-j|a/\zeta}
\end{equation}
is the covariance matrix. Each component of $\boldsymbol{\delta\mu}_\text{K}$ has zero mean and standard deviation $\sigma_\mu$, furthermore the length scale of the correlations is $\zeta$.}

\noteandras{The Cholesky decomposition of $\mathbf{\Sigma}$ has the form
\begin{equation}
	\mathbf{\Sigma} = \mathbf{L}\mathbf{L}^\intercal,
\end{equation} 
where $\mathbf{L}$ is a lower triangular matrix. We note that $\mathbf{\Sigma}$ is a real-valued symmetric positive-definite matrix. Let $\mathbf{Z}$ be an $N$-dimensional standard normal random vector. All components of $\mathbf{Z}$ are independent and each is a zero-mean unit-variance normally distributed random variable. Straighforward to see that $\boldsymbol{\delta\mu}_\text{K} = \mathbf{L}\mathbf{Z}$ follows the desired distribution with the covariance matrix described in Eq.~\eqref{eq:covmatrixappendix}. Thus to generate correlated random samples of on-site disorder, one can first generate uncorrelated samples (according to $\mathbf{Z}$), and then multiply them by the matrix $\mathbf{L}$.}

\bibliography{refs.bib}

\end{document}